\documentclass[reprint,amssymb, amsmath, aip,cha,twocolumn]{revtex4-2}
\usepackage{graphicx}
\usepackage[caption = false]{subfig}
\usepackage{color}
\usepackage{epsfig}
\usepackage{makeidx}
\usepackage{ifpdf}
\usepackage{url}%
\usepackage{float}
\usepackage{placeins}
\usepackage{subcaption}
\usepackage{bm}
\usepackage[colorlinks=true,linkcolor=blue]{hyperref}%
\usepackage{cleveref}
\expandafter\ifx\csname package@font\endcsname\relax\else
\expandafter\expandafter
\expandafter\usepackage
\expandafter\expandafter
\expandafter{\csname package@font\endcsname}%
\fi
\begin{document}
	\title{A Comprehensive Study on the Line Profiles and Stark Widths of Ionic Transitions from Laser Produced Aluminum Plasma }%
	\author{B R Geethika }%
	\email{geethika.br@ipr.res.in}
	\affiliation{Institute For Plasma Research, Bhat, Gandhinagar, Gujarat, 382428, India}%
	\affiliation{Homi Bhabha National Institute, Training School Complex, Anushaktinagar, Mumbai, 400094, India}%
	
	\author{Judhistir Shamal}
	\affiliation{Institute For Plasma Research, Bhat, Gandhinagar, Gujarat, 382428, India}%
	\affiliation{Homi Bhabha National Institute, Training School Complex, Anushaktinagar, Mumbai, 400094, India}%
	
	\author{Renjith Kumar R}
	\affiliation{Institute For Plasma Research, Bhat, Gandhinagar, Gujarat, 382428, India}%
	\affiliation{Homi Bhabha National Institute, Training School Complex, Anushaktinagar, Mumbai, 400094, India}%
	
	\author{Hem Chandra Joshi}
	\affiliation{Institute For Plasma Research, Bhat, Gandhinagar, Gujarat, 382428, India}%

	\author{Jinto Thomas}
	\email{jinto@ipr.res.in}
	\affiliation{Institute For Plasma Research, Bhat, Gandhinagar, Gujarat, 382428, India}%
	\affiliation{Homi Bhabha National Institute, Training School Complex, Anushaktinagar, Mumbai, 400094, India}%
	\date{\today}
	\begin{abstract}
		
		We present a systematic spectroscopic investigation of laser-produced aluminum plasma to address inconsistencies in Stark broadening parameters and establish a self-consistent reference datasets for electron density diagnostics. Optical line emissions of Al II and Al III in the visible wavelength range were recorded from plasmas having different electron densities and temperatures, however, with the same experimental configuration, only by varying the background pressure, spatial positions, and delay time. The Stark width parameter of Al III lines, which shows consistency across different earlier studies, is used for standardizing the Al II transition from the highest energy level, which is abundant in the emission spectra. This reference spectrum is then used to estimate the Stark parameters of other Al II transitions to obtain a self-consistent database for Al II transitions. This approach significantly reduces the uncertainty in the estimated plasma electron density using Stark parameters of multiple emission lines. We also report the spatial and temporal evolution of plasma density and Stark shift as well as asymmetry in spectral lines. This work addresses the uncertainty in Stark parameters of Al II transitions in the visible range through a unified approach in estimating these parameters simultaneously.
		
	\end{abstract}
	\maketitle
	\section{Introduction}\label{sec:intro}
	
	The optical emission from Laser Produced Plasma (LPP) carry vital information about the physical processes occurring within the plasma, making spectroscopic analysis a powerful diagnostic tool for investigating plasma parameters and behavior. 
	The underlying plasma dynamics modify the spectral line profiles, including line width, peak shifts, and line asymmetry\cite{grieim,CVEJIC201320,BENGOECHEA2005897,Stehle2000}. Numerous studies have estimated these parameters both experimentally and theoretically\cite{Konjevic,BURGER2016118,SAHALBRECHOT20141148,Alonso,Ortiz_2005}.
	Previously conducted studies on Stark parameters in various laser-produced plasmas, reveal that these effects are strongly correlated with the electron density but weakly on electron temperature\cite{Fleurier_1977,CIRISAN2014652,Kielkopf_2014}.

	The Stark shift and spectral line broadening arise from the perturbation of atomic energy levels by the electric micro fields of surrounding charged particles in the plasma, providing direct indicators of these internal electric fields\cite{grieim}. The magnitude and direction of the shift depend on perturbation of the particular transition with the local electric field, which is determined by the Stark sensitivity of the involved energy levels. By measuring line shifts we can infer the strength and the variation of these local fields, and learn how they change as the plasma expands and cools\cite{Djurovic,Gigosos2006}. Hence, studying the evolution of these parameters helps us understand both local conditions and the overall dynamics of the expanding plume\cite{grieim,Gigosos2006}.
	
	Further, the asymmetry in the line profile, on the other hand, originates from the non-uniform distribution and temporal variation of these electric micro-fields in the plasma. Fast-moving electrons mainly cause symmetric broadening through collisions, while slower ions generate quasi-static fields that vary spatially, resulting in unequal intensities for the two wings of the spectral line\cite{Gigosos2006,STAMBULCHIK20109}. Additional factors such as plasma electron density and temperature gradients and self-absorption can enhance or modify this asymmetry\cite{grieim,ELSHERBINI20051573,Hermann}. Therefore, studying the amount asymmetry also provides valuable information about the local electric environment and helps in understanding the microscopic field structure within the plasma.
	
	Accurate estimation of spectral parameters requires careful consideration of several prerequisites, particularly given the highly transient and spatially inhomogeneous nature of LPP\cite{HARILAL2018}. Since the plasma expands rapidly following laser ablation, measurements must be performed within a temporal window over which the plasma parameters remain approximately constant, thereby satisfying the quasi-stationary condition\cite{BENGOECHEA200669}. Similarly, spatially integrated measurements can distort the observed line profiles due to spatial gradients in plasma parameters along the line of sight\cite{Burger2019}. This can be corrected through Abel inversion, which recovers the localized emission profile assuming appropriate geometry of the expanding plume\cite{HARILAL2018}. Further, it is equally important to pay attention all other broadening mechanisms contributing to the observed line profile and must be accounted for. For example these include instrumental broadening, Doppler broadening, van der Waals broadening etc. in addition to the dominant Stark broadening.\cite{GORNUSHKIN19991207}. Once the contributions of these mechanisms are systematically accounted, the Stark broadening can be reliably attributed to electron density\cite{Alonso,Burger2019}. 
	
	Among different materials, aluminum (Al) is extensively studied  in laser plasma spectroscopy and laser ablation studies due to its simple atomic structure and well defined atomic emission lines. Previous investigations on Al lines have shown that their Stark widths and shifts scale with electron density as predicted by semi-classical impact theory, and the observed asymmetries are consistent with the combined influence of ionic quasi-static fields and self-absorption effects\cite{CIRISAN2014652,densal3,density704,density559}.
	For aluminum plasmas, Al II lines are particularly helpful because they are fairly intense and exists over a longer duration compared to the other atomic/ionic transitions\cite{density704,Konjevic,CIRISAN2014652}. 
	
	However, despite the numerous studies conducted on aluminum plasma, there is a considerable discrepancy in the experimentally reported values of Stark widths for aluminum line emissions. The reported Stark widths for Al II transitions in the visible wavelength range, which are extensively used, show variations as high as 100\%  \cite{CIRISAN2014652,densal3,density704,density559,Allen,heuschkel1973stark,Puri,BLAGOJEVIC20179,Konjevic} (ref- Table \ref{tab:stark_width_shift}). However, the reported value of Al III shows some coherence, where the variation is only within 10\%\cite{densal3}. The variations in reported values of Stark width may have dependency on experimental  systems like spatial resolution, spectral resolution, signal intensities etc, as most of this reported values are from different experiments performed by different experimental groups. One approach to minimize the spread in reported Stark width values is to record all relevant emission lines using the same experimental setup, so that instrumental contributions such as finite spectral resolution, spectral response, and temporal gating etc. remain constant across measurements.

	Though the estimation of plasma electron density using Stark parameters is a convenient method, the uncertainty of reported Stark parameters makes it difficult to ascertain the accuracy of the estimate. Hence, in this work, we attempt to reduce the reported inconsistencies in Stark width parameters of different aluminum lines in the visible wavelength range by measuring the Stark parameters of different lines (Al II and Al III) from the same experimental setup for a range of plasma parameters. We also focus on the spatio- temporal evolution of the spectral properties of Al II emission lines from laser-produced aluminum plasma under different background pressures. Further, the present study shows that the variations in spectral line width, shift, and asymmetry are correlated to the plasma parameters and ambiance. We believe that this study will help in having deeper understanding on the modifications of spectral properties in a high density plasma in addition to the evolution of plasma parameters and a comprehensive self consistent data for Stark parameters of Al II emissions in the visible wavelength range.

	\section{Experimental Set-up}\label{sec:setup}
	
	\begin{figure}[ht]
		\includegraphics[scale=0.27]{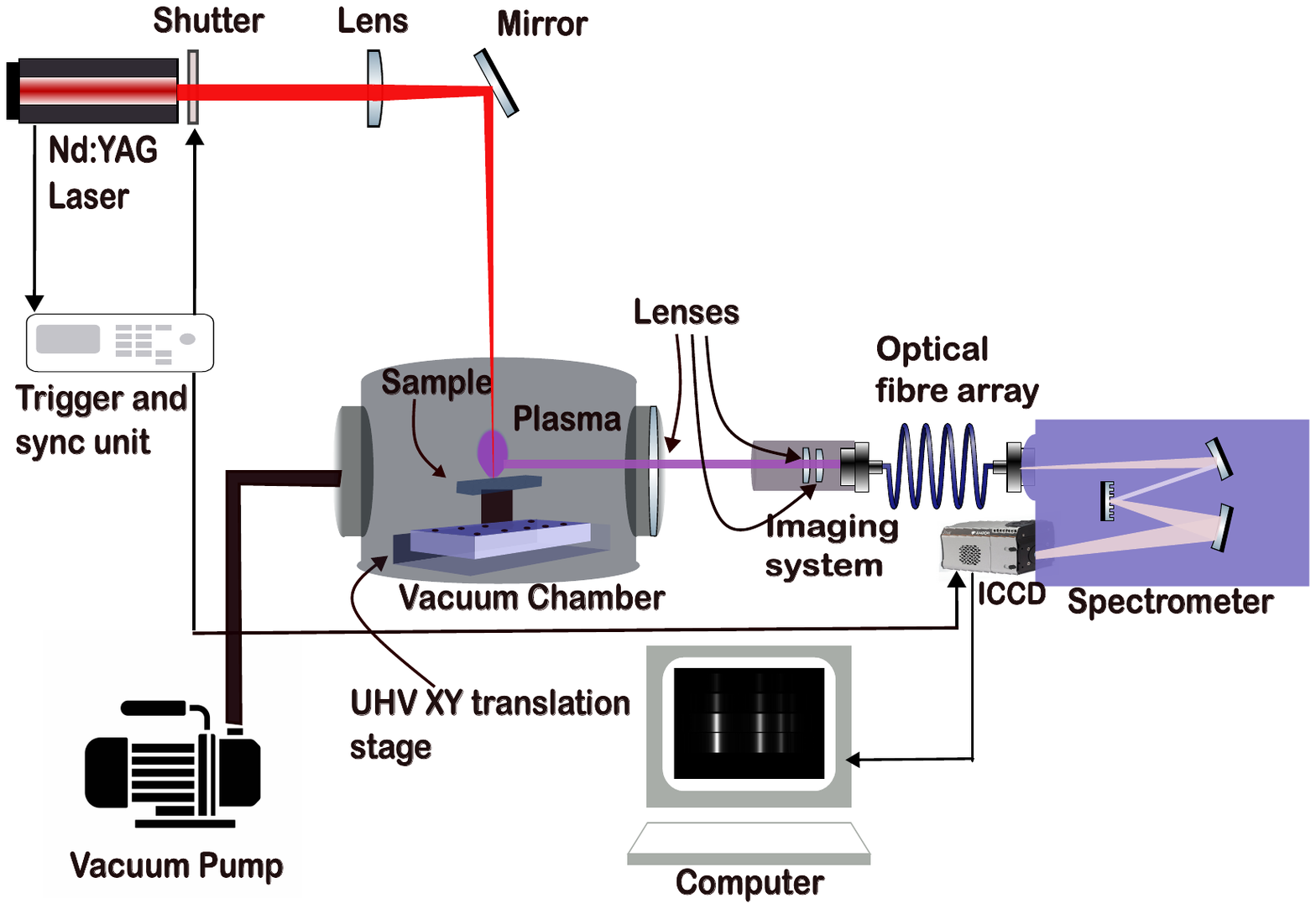}
		\caption{\label{fig:expsetup} A schematic diagram of experimental setup\cite{geethika_jaas}.  }
	\end{figure}
	
	
	Figure \ref{fig:expsetup} illustrates the schematic of the experimental configuration employed in this study. A polished aluminum sample with dimensions of 50 mm $\times$ 50 mm $\times$ 3 mm was positioned on a motorized translation stage within a stainless steel vacuum chamber. Prior to measurements, the chamber was pumped down to approximately $10^{-6}$ mbar using a turbo molecular pump. Argon was introduced as the ambient gas, with its pressure regulated through a precision needle valve.
	
	A Q-switched Nd:YAG laser operating at its fundamental wavelength of 1064 nm was used to ablate the sample. The laser delivered pulses of approximately 10 ns duration with an energy of 150 mJ per pulse. A 50 cm focal length plano-convex lens is used to loosely focus the beam onto the target producing a spot diameter of roughly 1 mm, which corresponds to a fluence of approximately 40 J/cm$^2$.
	
	Optical emission from the expanding plasma plume was captured through an in- house built imaging system and coupled to a one meter focal length Czerny-Turner spectrometer using a fiber optic bundle. Spectral acquisition was performed using an intensified charge-coupled device (ICCD) detector, enabling time-resolved measurements of Al II and Al III emission lines across the visible spectral region.
	
	The imaging system coupled with optical fiber provided spatial resolution of approximately 0.5 mm, which minimizes but does not eliminate line of sight integration effects.
	The temporal evolution of the spectral features was investigated by systematically varying the ICCD gate delay from 100 ns to 1500 ns relative to the laser pulse, with a gate width of 30 ns, ensuring that the selected time window satisfies the quasi-stationary condition. To reduce statistical fluctuations, each recorded spectrum represents an accumulation of 20 laser shots, with pulse-to-pulse intensity variations maintained below $\sim$5\%. The instrumental spectral resolution (0.06 nm) and profile (Lorentzian) were determined using a low-pressure calibration lamp. This instrumental broadening is subtracted from the measured line profiles to extract the Stark width. The Doppler broadening and the van der Waals broadening are much smaller than the instrumental broadening and therefore neglected in the present case.

\section{Results and Discussion}\label{sec:results}

\begin{figure}[h]
	\includegraphics[scale=0.3,trim = {0.7cm 0 0cm 0cm}, clip]{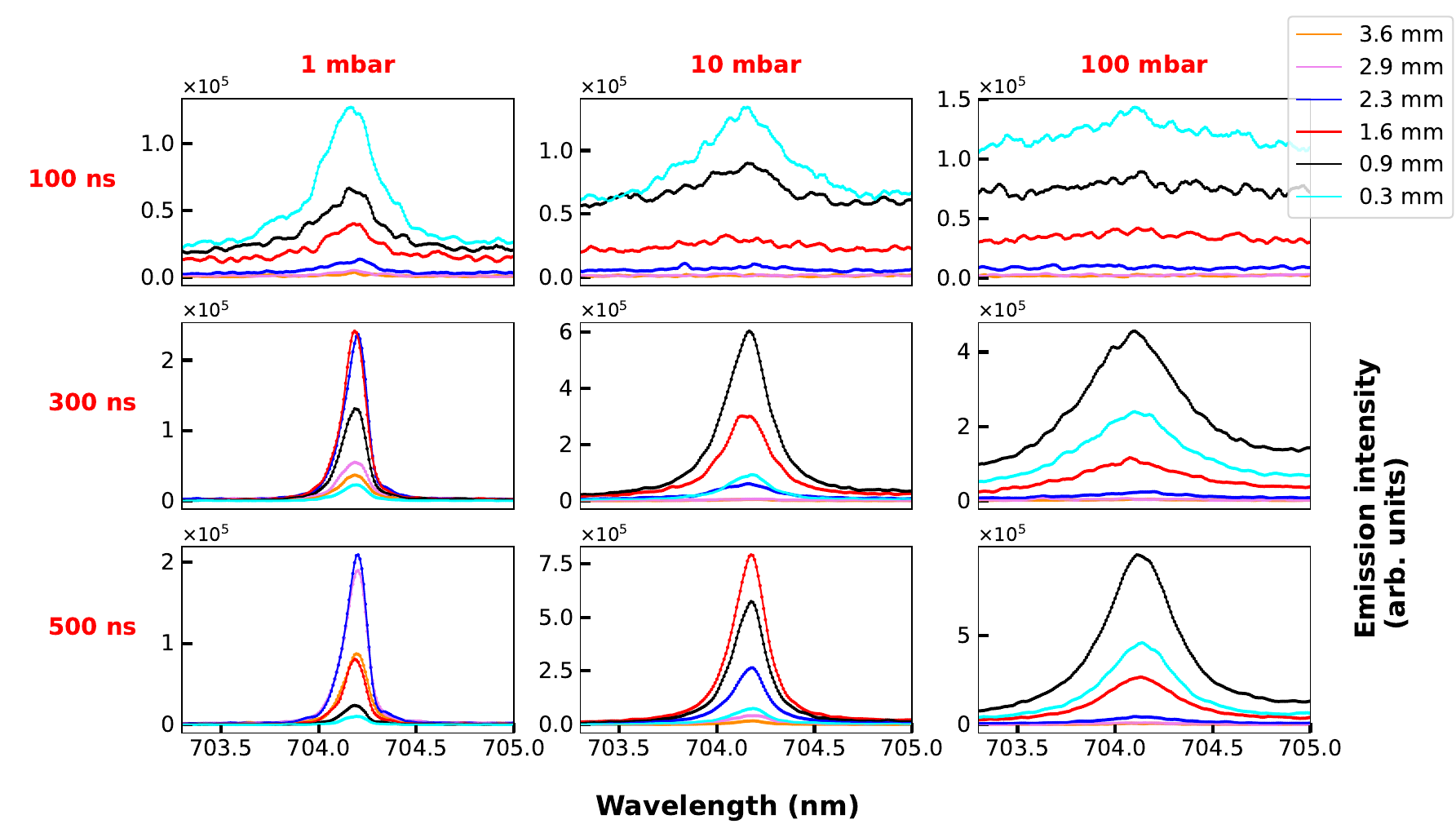}
	\caption{\label{fig:Spectra_3__lens} Spatio-temporal variation of various Al II emissions peaking at 704.21 nm at different background pressure ranging from 1 - 100 mbar.  }
\end{figure}


Figure~\ref{fig:Spectra_3__lens} shows the spatio-temporal evolution of aluminum plasma emission at 704.21 nm under three different ambient pressure conditions (1, 10, and 100 mbar) at three time delays (100, 300 and 500 ns). Different colors represent  different spatial positions from the target surface, ranging from 0.3 mm (cyan) to 3.6 mm (orange), illustrating the general plasma plume expansion dynamics.
The emission spectra demonstrate that the spectral properties are influenced by background pressure, spatial position and delay times. For instance, at lower pressures (1 mbar), the spectral lines starts appearing within 100 ns, whereas at higher pressures (10 and 100 mbar), the emission resembles a continuum or an extremely broadened line.

\begin{figure}[h]
	\includegraphics[scale=0.24,trim = {0.2cm 0 0cm 0cm}, clip]{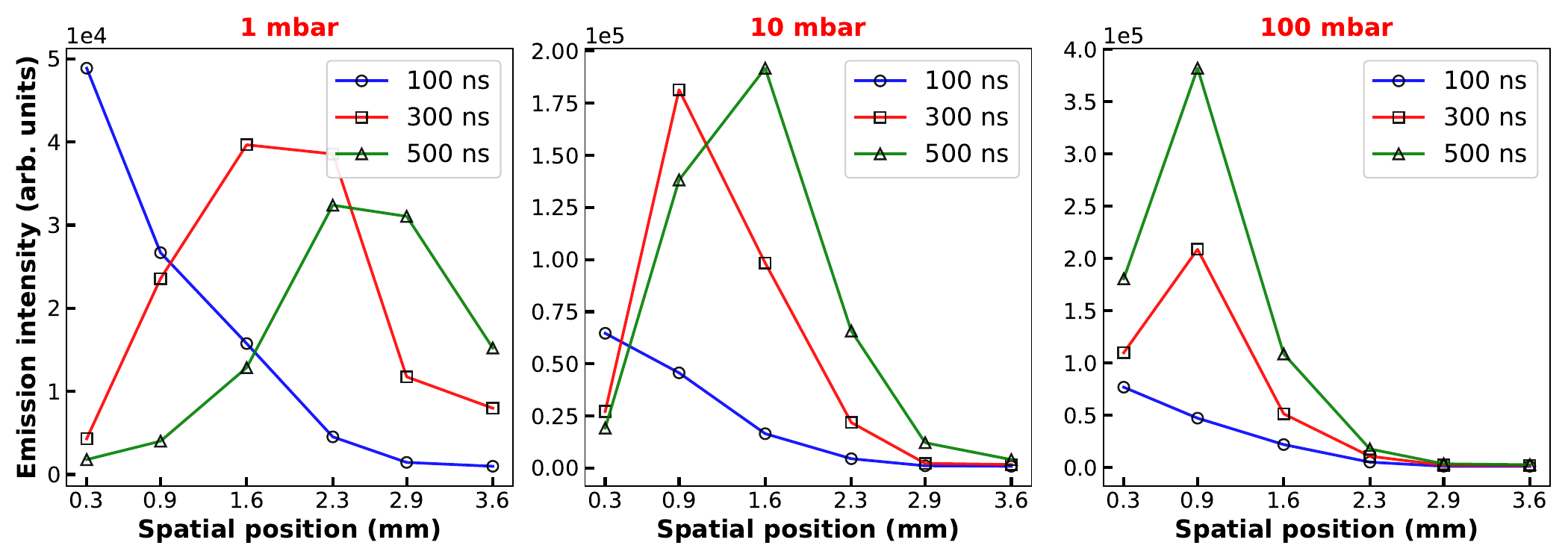}
	\caption{\label{fig:abs_int_704_var_pos_diffpress} Variation of integrated emission intensity of the Al I 704.21 nm line with spatial position at three background pressures (1, 10, and 100 mbar) and time delays (100, 300, and 500 ns). }
\end{figure}

To better understand the behaviour of emission intensity, the integrated intensity of 704.21 nm line is plotted against the spatial position for different delay times and background pressures as shown in figure \ref{fig:abs_int_704_var_pos_diffpress}. At lower background pressures (1 mbar), the plume continues to expand as time evolves, as evidenced by the spatial emission profiles, where the peak emission intensity shifts to longer distances from the target surface with increasing time delay. At 10 mbar, the expansion is comparatively slower, with the peak intensity position showing minimal shift over time. At 100 mbar, the plume is almost completely confined, with the emission intensity consistently peaking near the target surface at all time delays.
The systematic variation in spectral characteristics across the pressure-time parametric space illustrates the strong influence of ambient pressure on plume confinement, expansion velocity, and emission intensity evolution. 
Another notable observation from figure \ref{fig:Spectra_3__lens} is the systematic variation of spectral width (Stark effect) of the lines with background pressure and plume evolution time, which can be directly correlated to the plasma electron density. The figure indicates that this experimental configuration can yield substantial spectroscopic data from plasma with varying parameters, such as electron density and temperature, when recorded at different spatial locations, delay times and background pressures. A uniform method of documentation is employed for the purpose of acquiring comprehensive information regarding the spectral emission lines in question. This approach is undertaken to facilitate a detailed analysis of the aforementioned spectral lines in plasmas that possess similar parameters.

\begin{table}[htbp]
	\centering
	\caption{Previously reported values of experimentally observed Stark widths and Stark shifts for selected Al II spectral lines.}
	\label{tab:stark_width_shift}
	\resizebox{0.5\textwidth}{!}{
		\begin{tabular}{|c|c|c|c|}
			\hline
			\textbf{Wavelength} &
			\textbf{Spectral}  &
			\textbf{Stark Width} &
			\textbf{Stark Shift} \\
			\textbf{(nm)} &
			\textbf{Term}&
			\textbf{(FWHM) (\AA)} &
			\textbf{(\AA)} \\
			\hline
			466.31 & $^1$P$_1$ $\rightarrow$ $^1$D$_2$  & 1.09\cite{densal3}, 0.5\cite{Allen}, 2.6\cite{heuschkel1973stark} & -  \\
			559.33 &  $^3$D$_2$ $\rightarrow$ $^1$P$_1$ & 4.4\cite{densal3}, 3.8\cite{density559} &  - \\
			623.17 &  $^3$D$_2$ $\rightarrow$ $^3$P$_1$ & 3.44\cite{BLAGOJEVIC20179} & 1.46\cite{BLAGOJEVIC20179} \\
			624.34 & $^3$D$_3$ $\rightarrow$ $^3$P$_2$ & - & - \\
			704.21 & $^3$P$_2$ $\rightarrow$ $^3$S$_1$ & 0.95\cite{Puri}, 2\cite{Konjevic} & -0.68\cite{Puri}, -0.73\cite{Konjevic} \\
			705.66 & $^3$P$_1$ $\rightarrow$ $^3$S$_1$ & 0.92\cite{Puri}, 1.94\cite{Konjevic}, 1.74\cite{BLAGOJEVIC20179} & -0.7\cite{Puri}, -0.75\cite{Konjevic}, -0.8\cite{BLAGOJEVIC20179} \\
			706.36 & $^3$P$_0$ $\rightarrow$ $^3$S$_0$ & 0.93\cite{Puri}, 1.96\cite{Konjevic} & -0.67\cite{Puri}, -0.72\cite{Konjevic} \\

			\hline
		\end{tabular}
	}
\end{table}

\begin{table*}
	\centering
	\caption{Variation of mean electron density (calculated from Al~III lines) and electron excitation temperature along with Stark widths and impact factors of selected Al~II lines at different spatial locations and delay times for a background pressure of 100 mbar.}
	\label{tab:dens_temp_fwhm}
	\begin{tabular}{|c|c|c|c|c|c|c|c|c|c|}
		\hline
		\shortstack{Position \\ (mm)} &
		\shortstack{Time \\ (ns)} &
		\shortstack{Electron \\ Density \\ $N_e$ \\ ($\times10^{17}$ \\ cm$^{-3}$)} &
		\shortstack{Electron \\ Excitation \\ Temperature \\ $T_e$ \\ (eV)} &
		\shortstack{FWHM \\ 559.33 nm \\ (\AA)} &
		\shortstack{Stark impact \\ factor \\ for 559.33 nm \\ ($N_e = $\\$ 1\times10^{17}$ cm$^{-3}$ \\  (\AA)} &
		\shortstack{FWHM \\ 466.31 nm \\ (\AA)} &
		\shortstack{Stark impact \\ factor \\ for 466.31 nm\\ $N_e = $\\$ 1\times10^{17}$ cm$^{-3}$\\  (\AA)} &
		\shortstack{FWHM \\ 704.21 nm \\ (\AA)} &
		\shortstack{Stark impact \\ factor \\ for 704.21 nm\\ $N_e = $\\$ 1\times10^{17}$ cm$^{-3}$\\  (\AA)} \\
		\hline\hline
		0.5 & 300 & 3.26 $\pm$ 0.29 & 2.08 $\pm$ 0.05 & 18.3 $\pm$ 0.27 & 5.6 $\pm$ 0.58 & 4.4 $\pm$ 0.05 & 1.3 $\pm$ 0.13 & 6.7 $\pm$ 0.04 & 2.1 $\pm$ 0.20 \\
		0.5 & 500 & 1.78 $\pm$ 0.16 & 1.81 $\pm$ 0.11 & 10.2 $\pm$ 0.11 & 5.7 $\pm$ 0.57 & 2.9 $\pm$ 0.02 & 1.6 $\pm$ 0.15 & 4.4 $\pm$ 0.02 & 2.5 $\pm$ 0.24 \\
		1.5 & 300 & 3.00 $\pm$ 0.23 & 1.95 $\pm$ 0.07 & 17.8 $\pm$ 0.17 & 5.9 $\pm$ 0.51 & 4.6 $\pm$ 0.03 & 1.5 $\pm$ 0.12 & 8.3 $\pm$ 0.04 & 2.7 $\pm$ 0.22 \\
		1.5 & 500 & 1.85 $\pm$ 0.10 & 1.70 $\pm$ 0.10 & 10.9 $\pm$ 0.08 & 5.9 $\pm$ 0.36 & 3.1 $\pm$ 0.01 & 1.7 $\pm$ 0.10 & 4.7 $\pm$ 0.01 & 2.5 $\pm$ 0.14 \\
		2.5 & 300 & 3.62 $\pm$ 0.57 & 2.08 $\pm$ 0.10 & 19.4 $\pm$ 0.08 & 5.4 $\pm$ 0.87 & 5.3 $\pm$ 0.03 & 1.5 $\pm$ 0.24 & 8.5 $\pm$ 0.03 & 2.3 $\pm$ 0.37 \\
		2.5 & 500 & 1.90 $\pm$ 0.35 & 1.75 $\pm$ 0.07 & 9.9 $\pm$ 0.04 & 5.2 $\pm$ 0.98 & 2.9 $\pm$ 0.01 & 1.5 $\pm$ 0.28 & 4.6 $\pm$ 0.01 & 2.4 $\pm$ 0.45 \\
		3.5 & 300 & 4.00 $\pm$ 0.63 & 2.15 $\pm$ 0.05 & 21.3 $\pm$ 0.12 & 5.3 $\pm$ 0.86 & 4.9 $\pm$ 0.06 & 1.2 $\pm$ 0.20 & 8.8 $\pm$ 0.04 & 2.2 $\pm$ 0.36 \\
		3.5 & 500 & 1.93 $\pm$ 0.24 & 1.85 $\pm$ 0.07 & 10.4 $\pm$ 0.04 & 5.4 $\pm$ 0.69 & 3.1 $\pm$ 0.01 & 1.6 $\pm$ 0.20 & 5.0 $\pm$ 0.01 & 2.6 $\pm$ 0.33 \\
		\hline
	\end{tabular}
\end{table*}

Electron density in laser-produced plasma is normally determined from the Stark broadening of the spectral lines, the dominant line-broadening mechanism at high densities ($> 10^{16} cm^{-3}$) typical of these plasmas. 
Stark broadening for non- 
hydrogenic elements can be estimated from the following expression\cite{grieim}
\begin{equation}
	\Delta \lambda_{1/2}=2\omega\left(\frac{N_e}{10^{16}}\right)
	\label{Stark_eq_small}
\end{equation}

where $N_e (cm^{-3})$ is the electron density and $\omega$ is the electron broadening parameter (for a density of $1 \times 10^{16} cm^{-3}$), which weakly depends on the plasma temperature.

The electron excitation temperature can be estimated using the ratio of line intensities of successive ionic states\cite{grieim,densharilal}. The line intensity ratio method using successive ionization stages is advantageous over the Boltzmann plot method as it is more appropriate under non-ideal local thermodynamic equilibrium conditions and transient conditions as it involves the ionization potential\cite{grieim}. 
The dependence of the ratio of intensities of successive ionized states is given by equation \ref{Temp_int_R}\cite{grieim}

\begin{align}
	\frac{I'}{I} &=\frac{f'g'\lambda^3}{fg\lambda'^3}(4\pi^{3/2}a_0^3N_e)^{-1}\bigg(\frac{k_BT_e}{E_H}\bigg)^{3/2} \label{Temp_int_R} \\
	&\times exp\left(\frac{-(E'+E_{\infty}-E-\Delta E_{\infty})}{k_BT_e}\right) \notag		
\end{align}

where $\lambda, f, g, I $ and $ E $ are the wavelength, oscillator strength, statistical weight, line intensity, and upper level energy respectively for the lower ionic state. Similarly, $ \lambda', f', g', I' $ and $ E' $ are for higher ionic state. $E_{\infty}$ is the ionization energy of the lower ionic state and $\Delta E_{\infty}$ is the correction to the ionization energy. $E_H$ is the ionization energy of hydrogen atom, $a_0$ is the Bohr radius and $k_B$ is the Boltzmann constant. $N_e (m^{-3})$ and $T_e$ (Kelvin) are the density and temperature of electrons in the plasma. Equation \ref{Temp_int_R} can be used to estimate the electron temperature from the line intensity ratio of Al II and Al III lines if an accurate estimate of plasma density is available and by assuming Local Thermodynamic Equilibrium (LTE) which can be verified by McWhirter criterion\cite{ARAGON2008893}.

\begin{figure}[h]
	\includegraphics[scale=0.3,trim = {3.1cm 0 0cm 0.2cm}, clip]{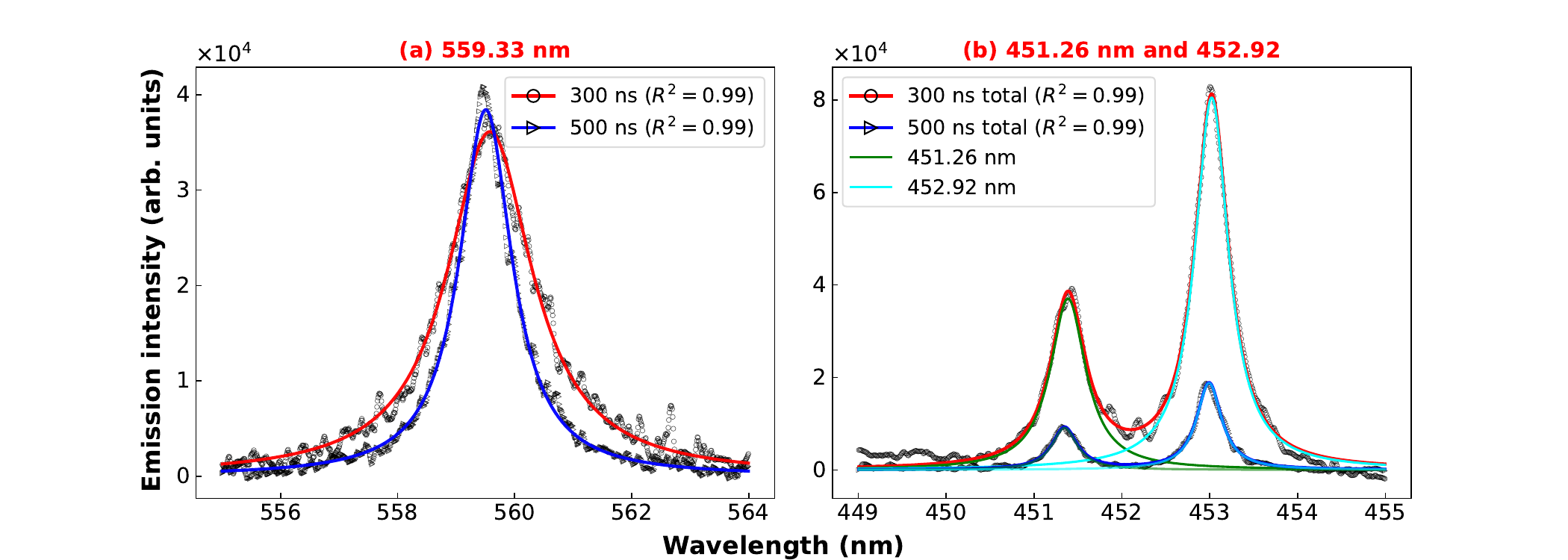}
	\caption{\label{fig:fit_559_452_100mbar_2plots_final} Typical emission spectra showing (a) the Al II line at 559.33 nm fitted with a single Lorentzian profile and (b) the Al III lines at 451.26 nm and 452.92 nm fitted with a two-peak Lorentzian profile. The $R^2$ values of the fits are included to indicate the goodness of fit. The Stark broadening widths extracted from the fits are included in Table \ref{tab:dens_temp_fwhm}.  }
\end{figure}

We have recorded the spectra of a number of Al II and Al III emissions in visible wavelength range for a range of  
background pressures, different locations within the plasma at different delay times for a particular laser fluence. This ensures that the spectra for a range of electron density ($1\times10^{16} cm^{-3} $ to $5\times10^{17} cm^{-3} $) and plasma electron temperature (1.5 to 2.5 eV) are recorded for all these lines. Figure \ref{fig:fit_559_452_100mbar_2plots_final} shows the acquired spectra of Al II  (559.33 nm) and Al III lines (451.26 nm and 452.92 nm) with Lorentzian fit and its corresponding $R^2$ value. It shows that the spectra fits well and the Stark parameters can be accurately estimated. Additionally, it may be noted that the Al III emission intensity drastically falls as the time increases(after 500 ns), as mentioned earlier.

The prolonged existence of Al II emission lines naturally makes them the preferred choice for estimating the plasma density using reported Stark widths. However, a comprehensive examination of the Stark width parameters reported in the literature\cite{densal3} reveals significant discrepancies among the reported widths. Table~\ref{tab:stark_width_shift} summaries the reported Stark parameters of Al II parameters. Indeed, the variations in the reported Stark widths for the prominent Al II line at 466.31 nm is substantial( vary by more than 100\% ).

As there is some consistency in the reported Stark widths of Al III lines\cite{densal3}. The electron density is first estimated using the Al III lines ( at  451.26, 452.92, 569.66 ,572.27 nm) reported by Dojić et al\cite{densal3}. Several Al III lines at fixed background pressure of 100 mbar are used for this and the mean value of density is estimated. The electron density calculated from different Al III lines are within a variation of 12\% further confirming the consistency in reported Stark width parameters for Al III lines. The electron temperature is also estimated from the line intensity ratios between different Al II and Al III lines using the density estimated and then averaged for a particular instance. The estimated plasma parameters are used to confirm the validity of LTE by McWhirter and Cristoforetti\cite{CRISTOFORETTI201086} criterion for the experiments. Table \ref{tab:dens_temp_fwhm} summarizes the electron density and temperature estimated as mentioned above for a particular position and time. The table also contains measured Stark widths for a few prominent Al II lines at corresponding location and time. Stark impact factor for electron density of $1 \times10^{17} cm^{-3}$ for the measured Al II lines are also provided in the table.

\begin{figure}[h]
	\includegraphics[scale=0.45,trim = {0cm 0 0 0.3cm}, clip]{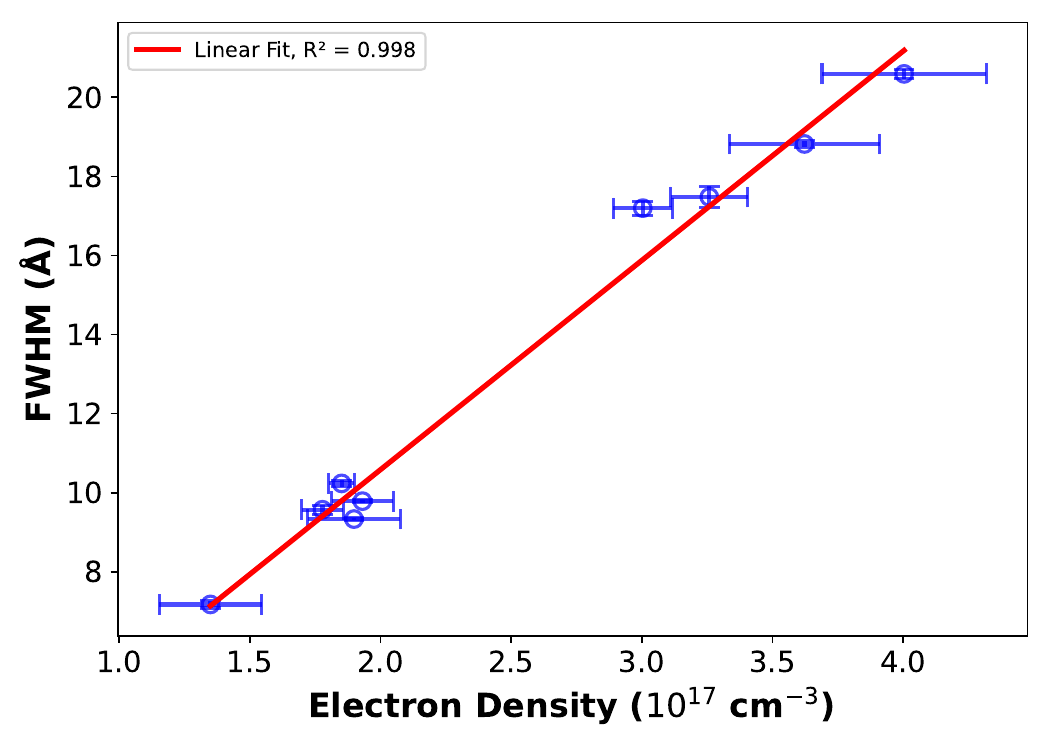}
	\caption{\label{fig:width559_densal3_latest} Variation of Stark width of Al II transition at 559.33 nm with the average electron density calculated from Al III lines. The error bar is the standard deviation of electron density calculated from different Al III lines.}
	\end{figure}

As the Al III emissions are absent for many instances, the density estimated from the Al III lines can be used to calibrate the Stark widths of the prominent Al II line 559.33 nm, which is mostly present for all instances and is expected to be less susceptible to self-absorption effects. The branching ratios of the Al II 704.21 nm/705.66 nm and 623.17 nm/624.34 nm transition pairs were examined to assess the possibility of self-absorption in these lines. The experimentally measured intensity ratios are found to be 1.6 $\pm$ 0.1 and 0.65 $\pm$ 0.08, which are in good agreement with the theoretical branching ratios of 1.68 and 0.61, respectively, calculated using transition probabilities from the NIST Atomic Spectra Database \cite{NIST_ASD}. In fact, these transitions originate from a lower-energy states compared to the 559.33 nm line further mitigates the concerns regarding self-absorption effects in the present measurements. 
Figure \ref{fig:width559_densal3_latest} shows the variation of Stark width of 559.33 nm line with the electron density estimated using the Stark width parameters of Al III lines\cite{densal3}. The linear fit for Stark width of 559.33 nm line with electron density gives very good $R^2$ value(0.997). It may be interesting to note that, despite significant temperature variation (as shown in table~\ref{tab:dens_temp_fwhm} 1.7 eV to 2.15 eV), the data fits well to a straight line with the Stark width of the 559.33 nm shows the minimal dependence of Stark width on electron temperature. The linear fit to the data yields a slope of $(5.29 \pm 0.08)\times10^{-17}$ $\AA / cm^{-3}$. The fit demonstrates that the Stark widths for 559.33 nm emission line exhibits a linear dependence with electron density and hence can be used for standardizing the other lines even in the absence of Al III lines during the experiments.

	\begin{figure}[h]
\includegraphics[scale=0.35,trim = {0cm 0 0 0.2cm}, clip]{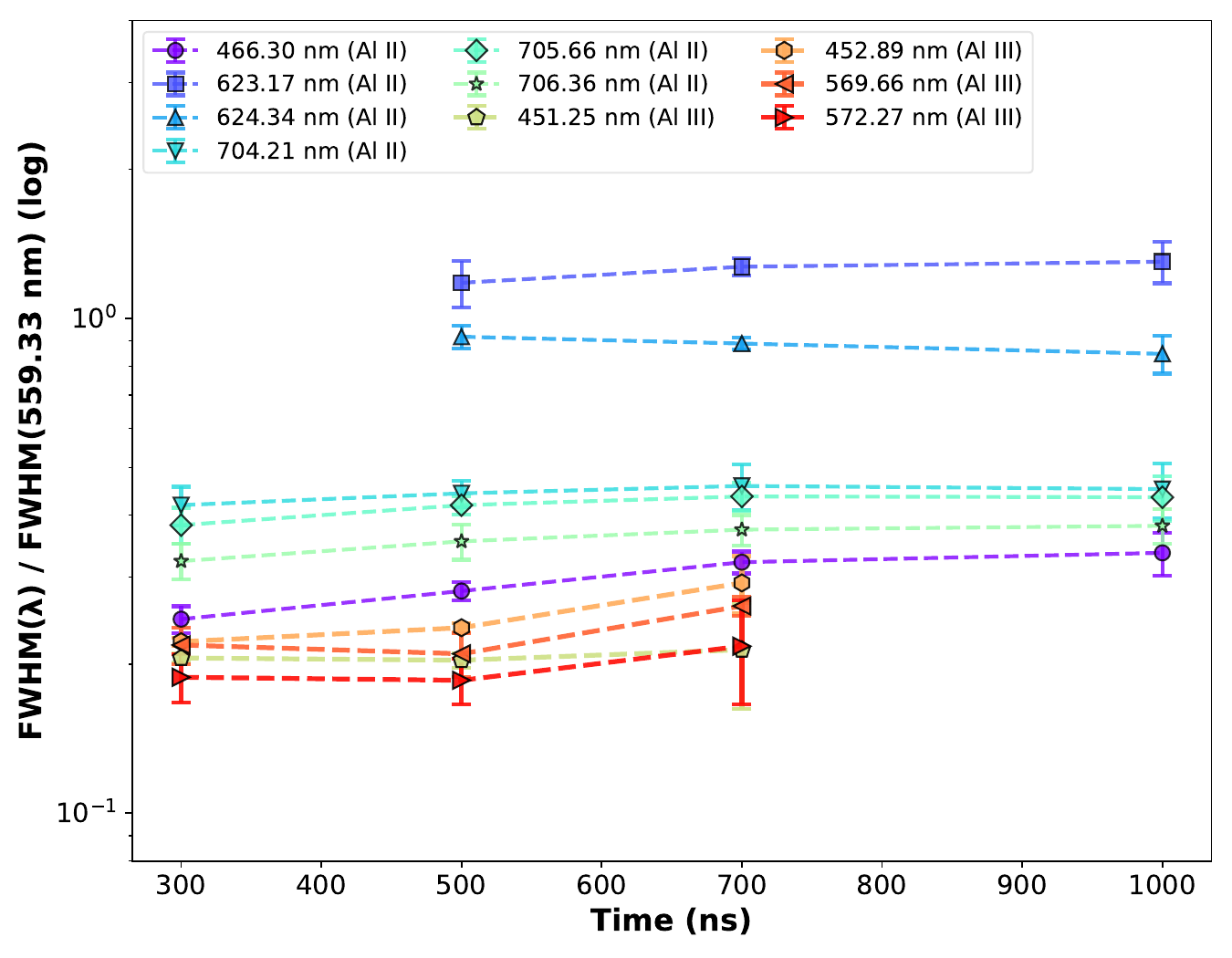}
\caption{\label{fig:fwhm_ratio_with_559_al2_al3_mod} Ratio of Stark widths of various transitions of Al II and Al III with 559.33 nm at different delay times. Error bar indicates the variations from different spatial locations.}
\end{figure}

To ensure the consistency in the measured spectral width with plasma electron density, we compared the spectral widths of various lines with the width of the 559.33 nm line recorded from the same experiment (by changing delay time, background pressure, and location).
Figure \ref{fig:fwhm_ratio_with_559_al2_al3_mod} shows the ratios between the widths of prominent Al III and Al II lines for different delay time for experiment performed at 100 mbar. The ratio of the widths from different spatial location is averaged and the deviation in ratio is expressed as the error bar. The figure clearly shows that the line width ratios of most lines remain constant with time.
The maximum variation in ratios observed for the Al lines are well within 15 \% indicating a good consistency in spectral widths.
\begin{figure}[h]
\includegraphics[scale=0.38,trim = {0.2cm 0 0 0.2cm}, clip]{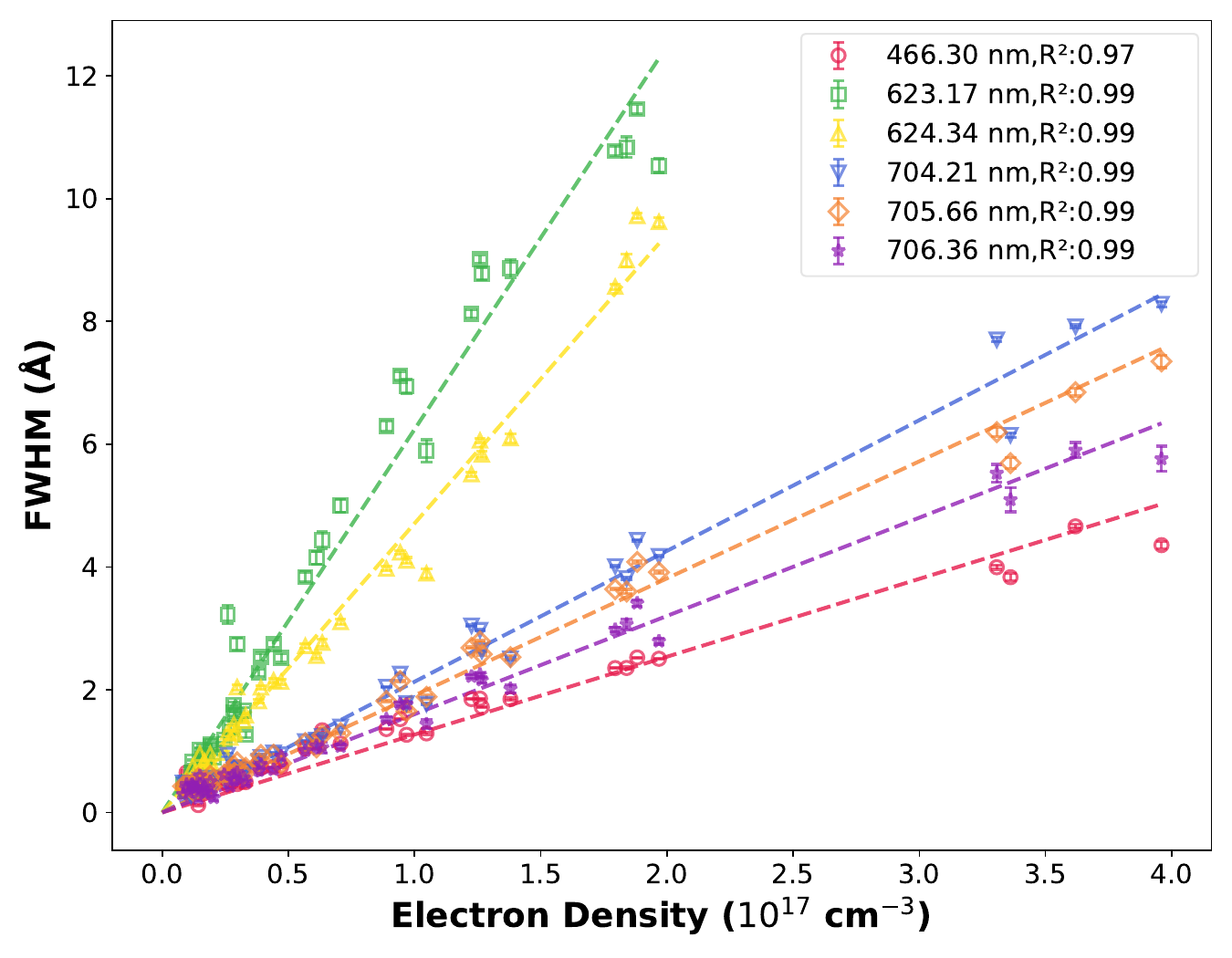}
\caption{\label{fig:fwhm_dens559_singleplot_7line_mod} Width of Al II transitions vs the electron density calculated from width of 559.33 nm  line. }
\end{figure}

As discussed in the previous section, we have determined the Stark impact factor for the Al II line at 559.33 nm to be $(5.29 \pm 0.08)\times10^{-17}$ $\AA/cm^{-3}$ from experiments for a range of plasma electron density and temperature. As stated earlier, this line is abundant in aluminum plasmas for the present parameter range than any of the Al III transitions. Hence, with this newly calculated Stark impact factor, we now determine the electron density at each spatial position and time delay for different background pressures. 
Similar to the validation of Stark impact factor of 559.33 nm using Al III lines, the widths of other Al II transitions are compared with the electron density determined from the 559.33 nm line, as shown in Figure \ref{fig:fwhm_dens559_singleplot_7line_mod}. Figure shows an excellent linear fit ($R^2 >0.97$)for the Stark width recorded for all the lines. This data can be safely used for the Stark widths of all these spectral lines that are recorded from the same experimental system and plasma parameters.

\begin{figure}[h!]
\includegraphics[scale=0.4,trim = {0cm 0 0 0.2cm}, clip]{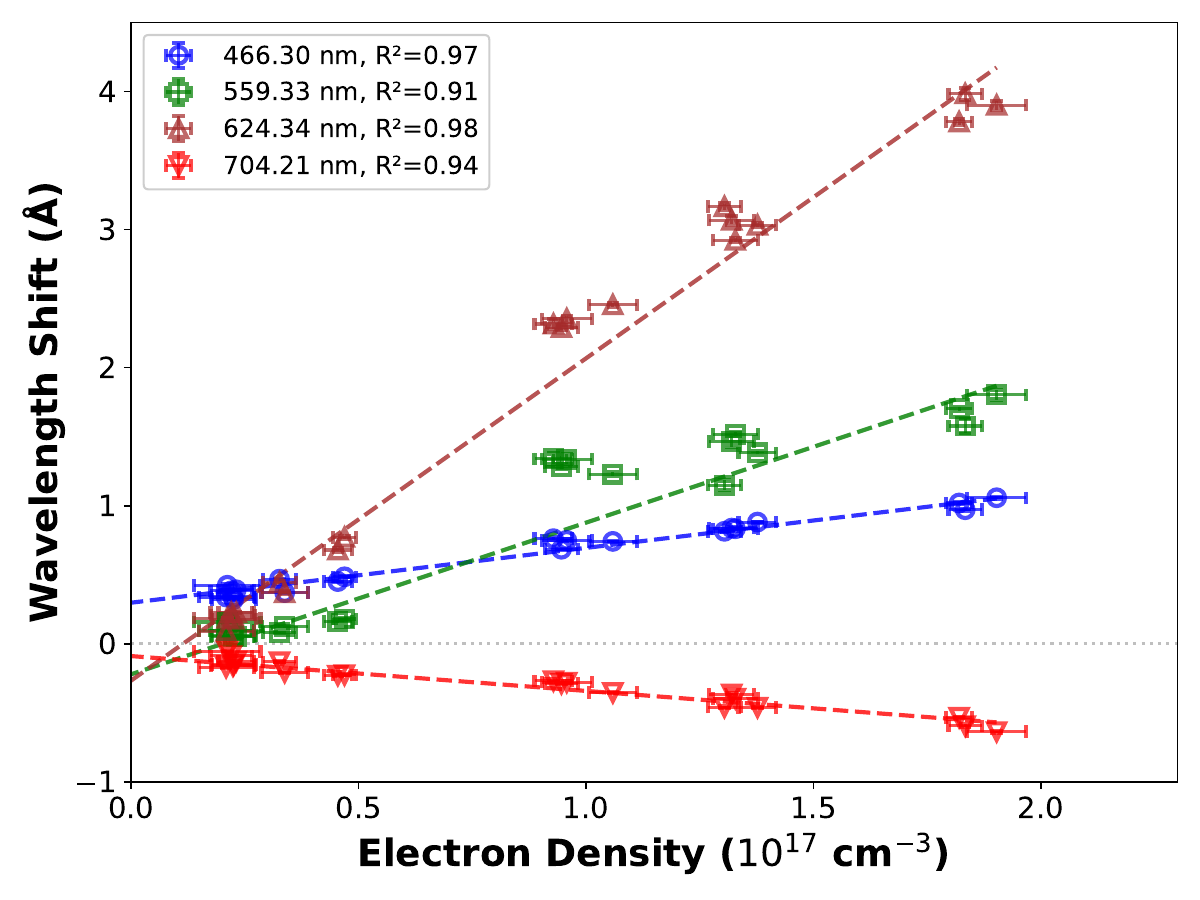}
\caption{\label{fig:density_vs_shift_allwave_final} Variation of Stark shift with plasma electron density for different Al II emissions. Error bar shows the standard deviation between the electron density calculated from different Al II lines. }
\end{figure}

\begin{figure*}[tbh]
\includegraphics[scale=0.6,trim = {0.2cm 0 0cm 0cm}, clip]{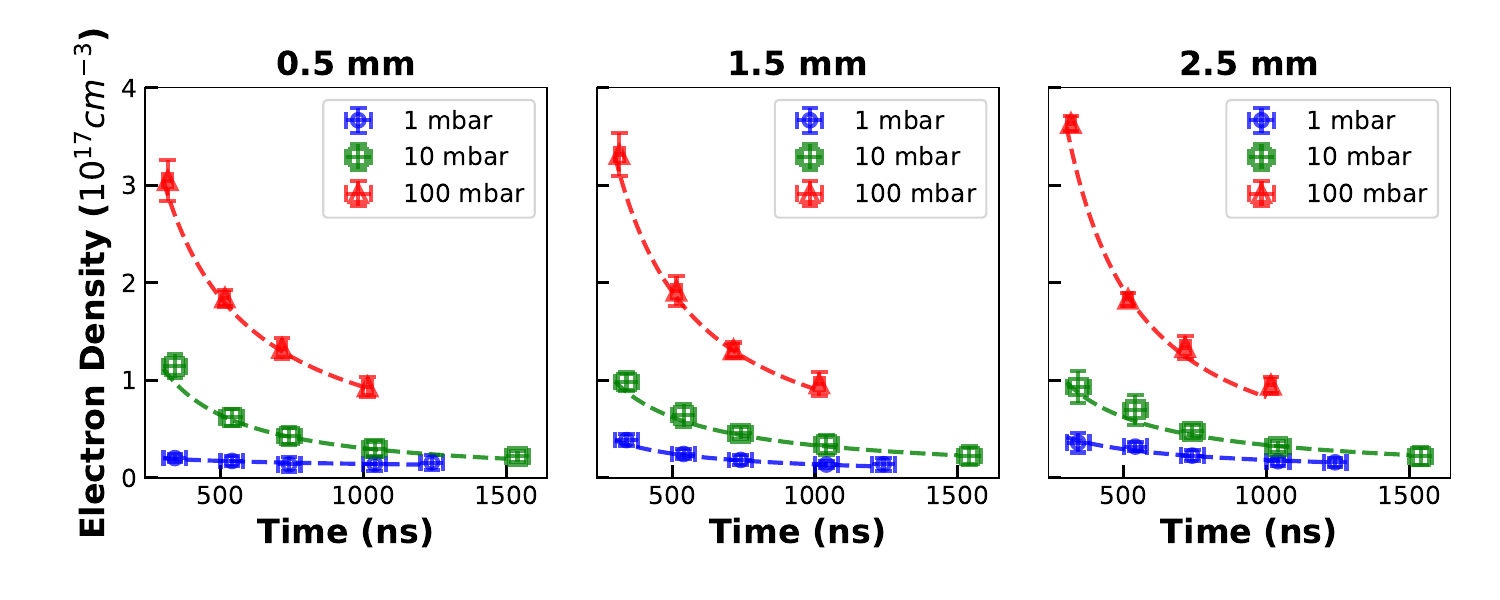}
\caption{\label{fig:dens_time_allpress_7linesal2_mod} Variation of plasma electron density at different positions with time for different background pressures. }
\end{figure*}

\begin{figure*}[tbh]

\begin{minipage}{1\textwidth}
\includegraphics[scale=0.45,trim = {0cm 0 0.5cm 0.6cm}, clip]{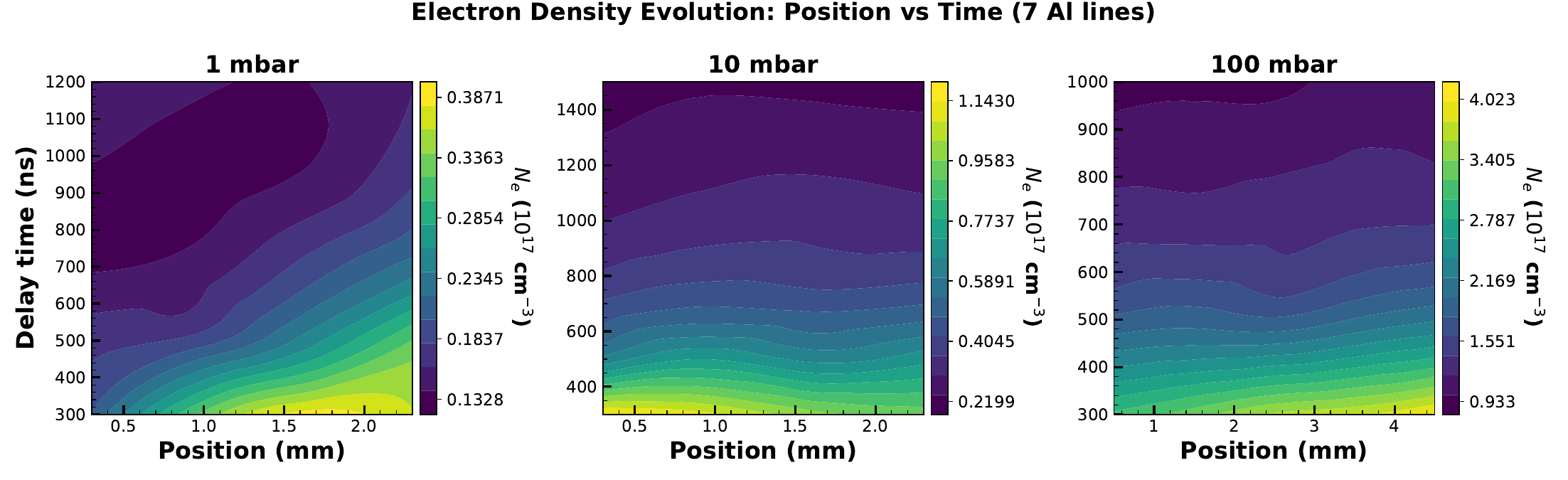}
\\ (a)   

\label{fig:density_diffpress_contour}
\end{minipage}

\begin{minipage}{1\textwidth}
\includegraphics[scale=0.45,trim = {0cm 0 0.5cm 0.1cm}, clip]{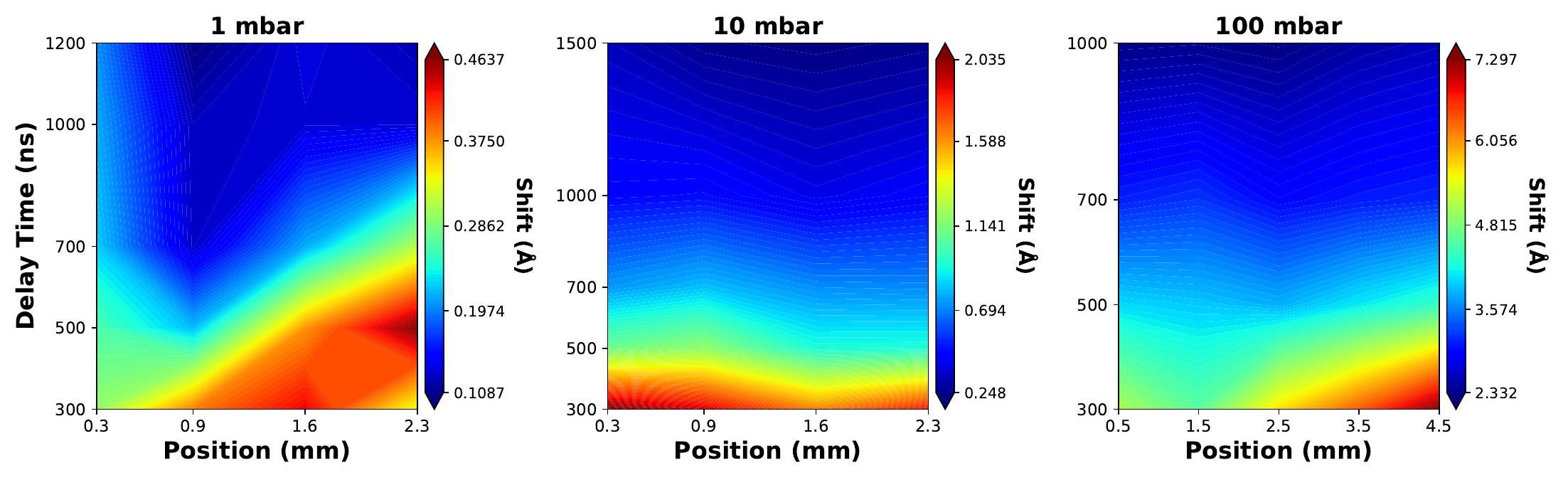}
\\ (b) 
\label{fig:shift_624_diffpress_contour}
\end{minipage}

\caption{\label{fig:dens_shift_cont} Spatio - temporal variation of (a) plasma electron density and (b) Stark shift (624.34 nm).}

\end{figure*}

Akin to the Stark width, we also estimate the Stark shift of these lines. The spectral profile of each transition is fitted with a Lorentzian function to estimate the shift from the original peak wavelength.
The shift in the peak wavelength is expected to be a function of Stark width (or Plasma electron density)\cite{Djurovic}.	Figure \ref{fig:density_vs_shift_allwave_final} shows the variation of shift in peak wavelength with plasma electron density for some of the transitions. The figure demonstrates that each transition exhibits a distinct slope, reflecting the line-specific Stark shift coefficients. Most of the transitions other than 704.21 show positive shifts (red shifts), while the 704.21 nm line shows a negative wavelength shift (blue shift) with increasing electron density\cite{density704}. This wavelength-dependent behavior of Stark shifts provides a means to validate electron density measurements and understand the atomic structure response to plasma electric fields. 
Additionally, the linear fit on the data is reasonably good with an $R^2$ value better than 0.91. It is interesting to note that the maximum offset observed among this lines (0.03 nm for 466.30 nm) is smaller than the spectrometer resolution.

\begin{table}[tbh]
\caption{Emission wavelength and spectral terms of the observed emission lines of Al II with corresponding Stark width and Stark shift at plasma electron density of $1 \times 10^{17}$ cm$^{-3}$ (Wavelengths from NIST database\cite{NIST_ASD}). The uncertainties in the reported Stark parameters represent the statistical uncertainty in the slope of the linear fit.}
\centering
\begin{tabular}{ | p{1.7cm} | p{2.2cm} | p{2cm} | p{2cm} | }
\hline
Wavelength (nm) & Spectral Terms & Stark Width ($\AA$) & Stark Shift ($\AA$) \\ \hline
466.30 & $^1$P$_{0}$ $\rightarrow$ $^1$D$_{0}$ & 1.27 $\pm$ 0.03 & 0.40 $\pm$ 0.02 \\ \hline
559.33 & $^1$D$_{2}$ $\rightarrow$ $^1$P$_{1}$ & 5.29 $\pm$ 0.08 & 1.12 $\pm$ 0.08 \\ \hline
623.17 & $^3$D$_{2}$ $\rightarrow$ $^3$P$_{1}$ & 6.24 $\pm$ 0.11 & 1.78 $\pm$ 0.07 \\ \hline
624.34 & $^3$D$_{3}$ $\rightarrow$ $^3$P$_{2}$ & 4.71 $\pm$ 0.05 & 2.34 $\pm$ 0.08 \\ \hline
704.21 & $^3$P$_{2}$ $\rightarrow$ $^3$S$_{1}$ & 2.13 $\pm$ 0.03 & $-$0.26 $\pm$ 0.02 \\ \hline
705.66 & $^3$P$_{1}$ $\rightarrow$ $^3$S$_{1}$ & 1.91 $\pm$ 0.02 & $-$0.16 $\pm$ 0.02 \\ \hline
706.36 & $^3$P$_{0}$ $\rightarrow$ $^3$S$_{1}$ & 1.60 $\pm$ 0.02 & $-$0.10 $\pm$ 0.02 \\ \hline
\end{tabular}
\label{table:stark_parameters}
\end{table}

The Stark parameters for all analyzed spectral lines are summarized in Table \ref{table:stark_parameters}. The Stark width and shift parameters listed in the table are deduced from the slope of a linear fit of the respective measured parameter (width and peak shift), respectively as a function of electron density. The uncertainties are derived from the covariance matrix of the linear fit. Our values, estimated using the standardized 559.33 nm line, show significant differences from values reported in the literature. 
These variations likely arise from multiple contributing factors. Instrumental factors such as finite spectral resolution, spectral response, and integration time may introduce systematic variations across different experimental setups. LPP is inherently inhomogeneous, exhibiting significant spatial and temporal gradients in both electron density and temperature across the plume. As a result, measurements integrate the emission along the line of sight and over a finite gate width, meaning that different effective plasma conditions are sampled across experiments, further contributing to the observed variations~\cite{HARILAL2018}. While line-of-sight and temporal averaging effects cannot be entirely eliminated in LPP diagnostics, in the present study their influence is minimized through the use of a narrow gate width and imaging system with suitable aperture stops. The Stark width parameters reported here are therefore self-consistent, having been determined from the same experimental setup under identical plasma conditions.

Using the estimated Stark width parameters of all the spectral lines, the plasma electron density for different spatial locations and delay times are estimated.
Figure~\ref{fig:dens_time_allpress_7linesal2_mod} shows the temporal evolution of the electron density at different background pressures and spatial locations.
The standard deviation is significantly reduced ( less than 15\%) here compared to an estimation of plasma density using the earlier reported Stark width parameters taken from  different experimental works (None of the earlier studies reported the values for all these lines).  The results exhibit the expected temporal behavior, with the electron density following an approximate power-law decay 
with time, and a systematic increase in electron density with increasing background pressure.


It may be interesting to see the spatio- temporal evolution of plasma density in the present laser produced plasma experiment with different background pressures.
Figure \ref{fig:dens_shift_cont} presents the spatio-temporal evolution of (a) plasma electron density (b) wavelength shift 624.34 nm line (has maximum shift)
under varying background pressures. 
At low pressures (1 and 10 mbar),  emission line at 624.34 nm exhibits negligible wavelength shifts across all spatial positions and delay times similar to the trend in plasma electron density. In contrast, at 100 mbar background pressure, a pronounced wavelength shifts emerge with distinct spatio-temporal gradients.  This pressure-dependent behavior is attributed to the Stark effect, wherein increased background pressure enhances plasma electron density and stronger electric field interactions with the radiating ions\cite{grieim,Zehra_Bashir_Hassan_Ahmed_Akram_Hayat_2017,Kautz2020,Konjevic}.The electron density and the Stark shift decrease with increasing time delay, as expected from plume expansion and the associated reduction in electron density. However, a spatial dependence is observed for both electron density and peak shift, which likely arises from pressure-dependent plume dynamics, with free expansion at low pressures and increasing confinement at higher pressures \cite{Harilal2003}. At pressures as high as 100 mbar, the interaction of plasma with background increases significantly. Such interaction can results in localized density enhancements\cite{Harilal2003} as can be seen in the figure. However, this requires more experiments to to confirm. Additionally, line-of-sight integration effects in the absence of Abel inversion may slightly contribute to the observed spatial variations\cite{HARILAL2018}.

Another interesting observation from the spectroscopic study is the observation of asymmetry in line profile.
Asymmetry in line profiles can arise from multiple mechanisms, including quadrupole interactions between radiating ions and perturbing ions \cite{joyce_asym}, ion micro-field inhomogeneity \cite{grieim,PhysRevE_jinto}, self-absorption \cite{grieim}, ion dynamics arising from the violation of the quasi-static approximation \cite{ion_dynamics}, spatial gradients of electron density within the plasma \cite{Hermann2017}, etc.

\begin{figure}[h]
\centering

\begin{minipage}{0.23\textwidth}
\centering
\includegraphics[width=1.05\linewidth,trim=0cm 0 0 1.3cm,clip]{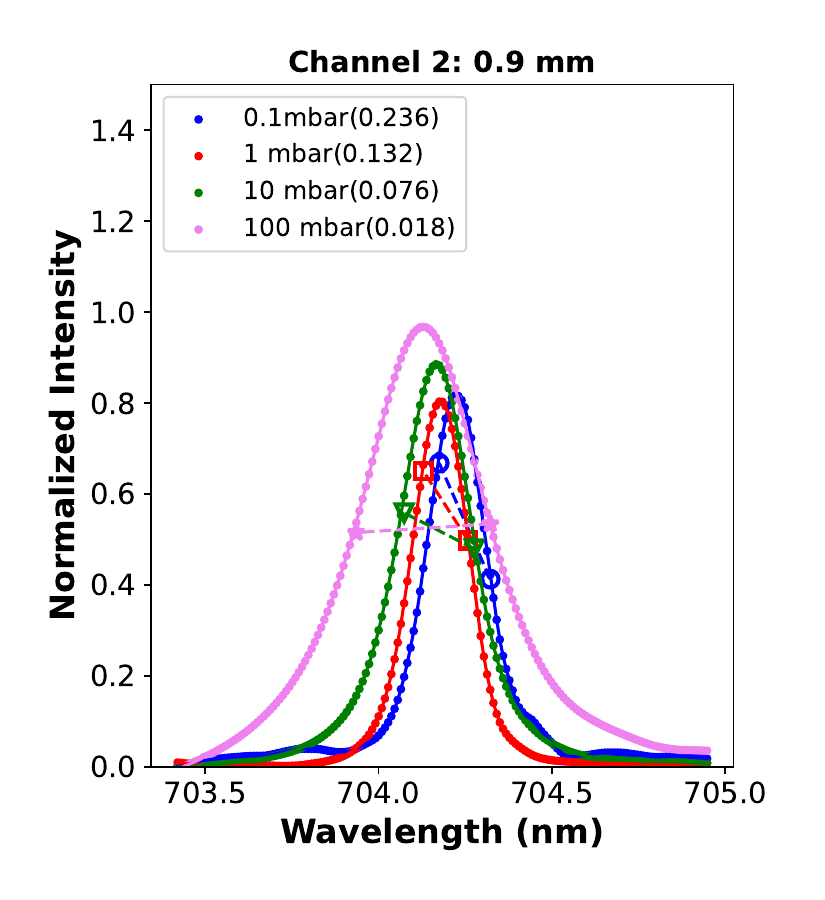}
\caption*{(a)}
\end{minipage}
\begin{minipage}{0.23\textwidth}
\centering
\includegraphics[width=0.96\linewidth,trim=0cm 0 0 1.3cm,clip]{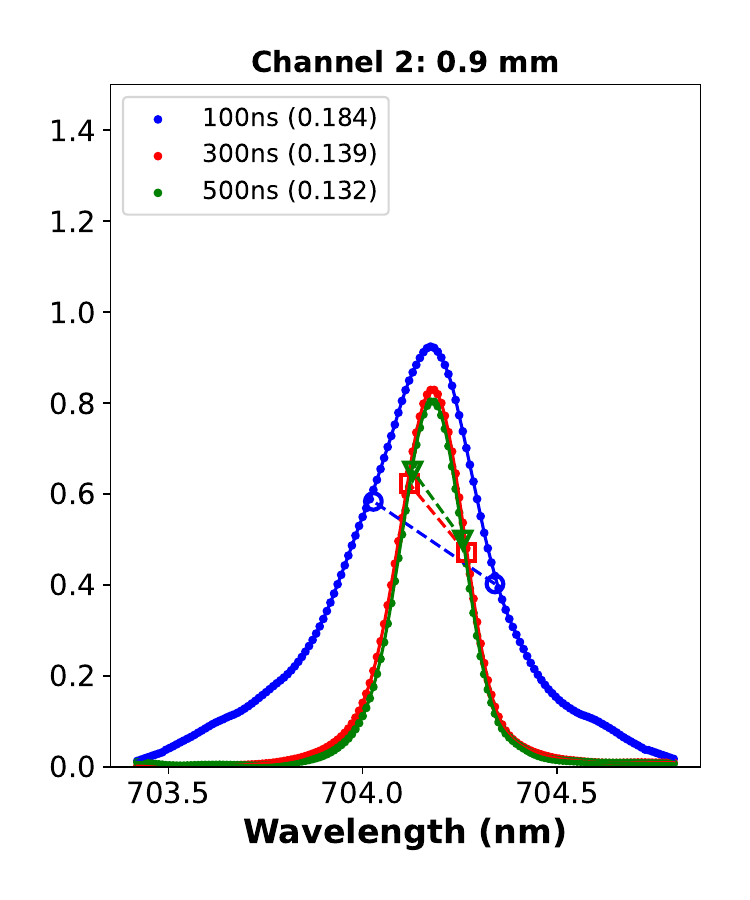}
\caption*{(b)}
\end{minipage}

\caption{Spectra of Al II line (704.21 nm) at 1 mm (a) varying background pressure at 500 ns (b) varying delay time at 1 mbar. The points on the each peak corresponds to the intensities at +$\delta \lambda$ and - $\delta \lambda$.}
\label{fig:two_figures}
\end{figure}

Here, the estimation of spectral asymmetry ($A_s$) for a given transition is estimated using the formula~\cite{PhysRevE_jinto}:
\begin{equation}
A_s(\delta\lambda) = \frac{I_R - I_B}{I_R + I_B},
\label{eq:asymmetry}
\end{equation}
where $\delta\lambda = FWHM/2$, $I_R$ and $I_B$ are the line intensities at wavelength separations of $+\delta\lambda$ and $-\delta\lambda$ from the line center, respectively. 
It is observed that the shift of peak wavelength and asymmetry are specific to transitions where some transitions shows a substantial shift and asymmetry  for the same plasma parameters\cite{Gigosos2006,grieim,Konjevic}.
Figure~\ref{fig:two_figures}(a) shows the asymmetry observed in the Al~II 704.21~nm line at different background pressures, measured at a distance of 1~mm from the target and at a delay of 500~ns after the laser pulse. The asymmetry remains relatively small at higher background pressures and increases as the background pressure decreases. This behavior may be attributed to the ion micro field effects or to breakdown of the quasi-static ion approximation at lower pressures, where plasma expansion and ion dynamics become more significant. However, a detailed analysis of these effects is to be further explored for a quantitative interpretation.  

Figure~\ref{fig:two_figures}(b) shows the temporal evolution of the spectral asymmetry at a background pressure of 1~mbar and a distance of 1~mm from the target. The asymmetry decreases with increasing delay time at a particular background pressure, which is consistent with the expected reduction in electron density as the plasma expands and cools. Consequently, the net electric micro field produced by the surrounding ions decreases, leading to a more symmetric line profile.

\section{conclusion}
\label{sec:conclusion}

In this study of laser-produced aluminum plasma, we address the inconsistencies in the reported Stark width parameters for Al II lines, which are likely arising from the differences in instrumental spectral resolutions, spectral response calibration, line profile fitting methodologies and spatiotemporal variations in plasma parameters inherent to LPPs.
The plasma electron density is varied by more than one order of magnitude to find the linear relation of Stark width parameters with electron density, and we also observed that the Stark width parameters are not highly sensitive to temperature variations of a few eV. Using the reported Al III Stark width parameters as a basis for estimation of plasma electron density, a number of prominent Al II transitions are bench marked for the Stark width parameters.

With the estimated Stark width parameters, the spatio-temporal evolution of plasma density is studied, and it is found that the variations in estimated electron density from different Al II lines are significantly less in comparison to the estimates using the earlier reported values of Stark widths from different literature sources. Further, the background pressure plays a critical role in the plasma density, Stark shift and asymmetric nature of lines. The plasma density and Stark effects increases with the background pressure. However, the spectral asymmetry of the 704.21 nm line is showing decrease in asymmetry with background pressure, indicating the role of ion dynamics in the observed asymmetry. The asymmetry also decreases with delay time.

\section{References}

%


\begin{thebibliography}{40}%
\makeatletter
\providecommand \@ifxundefined [1]{%
 \@ifx{#1\undefined}
}%
\providecommand \@ifnum [1]{%
 \ifnum #1\expandafter \@firstoftwo
 \else \expandafter \@secondoftwo
 \fi
}%
\providecommand \@ifx [1]{%
 \ifx #1\expandafter \@firstoftwo
 \else \expandafter \@secondoftwo
 \fi
}%
\providecommand \natexlab [1]{#1}%
\providecommand \enquote  [1]{``#1''}%
\providecommand \bibnamefont  [1]{#1}%
\providecommand \bibfnamefont [1]{#1}%
\providecommand \citenamefont [1]{#1}%
\providecommand \href@noop [0]{\@secondoftwo}%
\providecommand \href [0]{\begingroup \@sanitize@url \@href}%
\providecommand \@href[1]{\@@startlink{#1}\@@href}%
\providecommand \@@href[1]{\endgroup#1\@@endlink}%
\providecommand \@sanitize@url [0]{\catcode `\\12\catcode `\$12\catcode
  `\&12\catcode `\#12\catcode `\^12\catcode `\_12\catcode `\%12\relax}%
\providecommand \@@startlink[1]{}%
\providecommand \@@endlink[0]{}%
\providecommand \url  [0]{\begingroup\@sanitize@url \@url }%
\providecommand \@url [1]{\endgroup\@href {#1}{\urlprefix }}%
\providecommand \urlprefix  [0]{URL }%
\providecommand \Eprint [0]{\href }%
\providecommand \doibase [0]{https://doi.org/}%
\providecommand \selectlanguage [0]{\@gobble}%
\providecommand \bibinfo  [0]{\@secondoftwo}%
\providecommand \bibfield  [0]{\@secondoftwo}%
\providecommand \translation [1]{[#1]}%
\providecommand \BibitemOpen [0]{}%
\providecommand \bibitemStop [0]{}%
\providecommand \bibitemNoStop [0]{.\EOS\space}%
\providecommand \EOS [0]{\spacefactor3000\relax}%
\providecommand \BibitemShut  [1]{\csname bibitem#1\endcsname}%
\let\auto@bib@innerbib\@empty
\bibitem [{\citenamefont {Griem}(1964)}]{grieim}%
  \BibitemOpen
  \bibfield  {author} {\bibinfo {author} {\bibfnamefont {H.}~\bibnamefont
  {Griem}},\ }\href@noop {} {\emph {\bibinfo {title} {Plasma Spectroscopy}}}\
  (\bibinfo  {publisher} {McGraw-Hill},\ \bibinfo {year} {1964})\BibitemShut
  {NoStop}%
\bibitem [{\citenamefont {Cvejić}\ \emph {et~al.}(2013)\citenamefont
  {Cvejić}, \citenamefont {Gavrilović}, \citenamefont {Jovićević},\ and\
  \citenamefont {Konjević}}]{CVEJIC201320}%
  \BibitemOpen
  \bibfield  {author} {\bibinfo {author} {\bibfnamefont {M.}~\bibnamefont
  {Cvejić}}, \bibinfo {author} {\bibfnamefont {M.}~\bibnamefont
  {Gavrilović}}, \bibinfo {author} {\bibfnamefont {S.}~\bibnamefont
  {Jovićević}},\ and\ \bibinfo {author} {\bibfnamefont {N.}~\bibnamefont
  {Konjević}},\ }\bibfield  {title} {\enquote {\bibinfo {title} {{Stark}
  broadening of {Mg~I} and {Mg~II} spectral lines and {Debye} shielding effect
  in laser induced plasma},}\ }\href
  {https://doi.org/https://doi.org/10.1016/j.sab.2013.03.011} {\bibfield
  {journal} {\bibinfo  {journal} {Spectrochimica Acta Part B: Atomic
  Spectroscopy}\ }\textbf {\bibinfo {volume} {85}},\ \bibinfo {pages} {20--33}
  (\bibinfo {year} {2013})}\BibitemShut {NoStop}%
\bibitem [{\citenamefont {Bengoechea}, \citenamefont {Arag{\'o}n},\ and\
  \citenamefont {Aguilera}(2005)}]{BENGOECHEA2005897}%
  \BibitemOpen
  \bibfield  {author} {\bibinfo {author} {\bibfnamefont {J.}~\bibnamefont
  {Bengoechea}}, \bibinfo {author} {\bibfnamefont {C.}~\bibnamefont
  {Arag{\'o}n}},\ and\ \bibinfo {author} {\bibfnamefont {J.}~\bibnamefont
  {Aguilera}},\ }\bibfield  {title} {\enquote {\bibinfo {title} {Asymmetric
  {Stark} broadening of the {Fe~I} 538.34 nm emission line in a laser induced
  plasma},}\ }\href {https://doi.org/https://doi.org/10.1016/j.sab.2005.05.019}
  {\bibfield  {journal} {\bibinfo  {journal} {Spectrochimica Acta Part B:
  Atomic Spectroscopy}\ }\textbf {\bibinfo {volume} {60}},\ \bibinfo {pages}
  {897--904} (\bibinfo {year} {2005})},\ \bibinfo {note} {laser Induced Plasma
  Spectroscopy and Applications (LIBS 2004) Third International
  Conference}\BibitemShut {NoStop}%
\bibitem [{\citenamefont {Stehl{\'e}}, \citenamefont {Gilles},\ and\
  \citenamefont {Demura}(2000)}]{Stehle2000}%
  \BibitemOpen
  \bibfield  {author} {\bibinfo {author} {\bibfnamefont {C.}~\bibnamefont
  {Stehl{\'e}}}, \bibinfo {author} {\bibfnamefont {D.}~\bibnamefont {Gilles}},\
  and\ \bibinfo {author} {\bibfnamefont {A.~V.}\ \bibnamefont {Demura}},\
  }\bibfield  {title} {\enquote {\bibinfo {title} {Asymmetry of {Stark}
  profiles},}\ }\href {https://doi.org/10.1007/s100530070032} {\bibfield
  {journal} {\bibinfo  {journal} {The European Physical Journal D - Atomic,
  Molecular, Optical and Plasma Physics}\ }\textbf {\bibinfo {volume} {12}},\
  \bibinfo {pages} {355--367} (\bibinfo {year} {2000})}\BibitemShut {NoStop}%
\bibitem [{\citenamefont {Konjevi{\'c}}\ \emph {et~al.}(2002)\citenamefont
  {Konjevi{\'c}}, \citenamefont {Lesage}, \citenamefont {Fuhr},\ and\
  \citenamefont {Wiese}}]{Konjevic}%
  \BibitemOpen
  \bibfield  {author} {\bibinfo {author} {\bibfnamefont {N.}~\bibnamefont
  {Konjevi{\'c}}}, \bibinfo {author} {\bibfnamefont {A.}~\bibnamefont
  {Lesage}}, \bibinfo {author} {\bibfnamefont {J.~R.}\ \bibnamefont {Fuhr}},\
  and\ \bibinfo {author} {\bibfnamefont {W.~L.}\ \bibnamefont {Wiese}},\
  }\bibfield  {title} {\enquote {\bibinfo {title} {Experimental {Stark} widths
  and shifts for spectral lines of neutral and ionized atoms ({A} critical
  review of selected data for the period 1989 through 2000)},}\ }\href
  {https://doi.org/10.1063/1.1486456} {\bibfield  {journal} {\bibinfo
  {journal} {Journal of Physical and Chemical Reference Data}\ }\textbf
  {\bibinfo {volume} {31}},\ \bibinfo {pages} {819--927} (\bibinfo {year}
  {2002})}\BibitemShut {NoStop}%
\bibitem [{\citenamefont {Burger}\ and\ \citenamefont
  {Hermann}(2016)}]{BURGER2016118}%
  \BibitemOpen
  \bibfield  {author} {\bibinfo {author} {\bibfnamefont {M.}~\bibnamefont
  {Burger}}\ and\ \bibinfo {author} {\bibfnamefont {J.}~\bibnamefont
  {Hermann}},\ }\bibfield  {title} {\enquote {\bibinfo {title} {{Stark}
  broadening measurements in plasmas produced by laser ablation of hydrogen
  containing compounds},}\ }\href
  {https://doi.org/https://doi.org/10.1016/j.sab.2016.06.005} {\bibfield
  {journal} {\bibinfo  {journal} {Spectrochimica Acta Part B: Atomic
  Spectroscopy}\ }\textbf {\bibinfo {volume} {122}},\ \bibinfo {pages}
  {118--126} (\bibinfo {year} {2016})}\BibitemShut {NoStop}%
\bibitem [{\citenamefont {Sahal-Br{\'e}chot}\ \emph {et~al.}(2014)\citenamefont
  {Sahal-Br{\'e}chot}, \citenamefont {Dimitrijević}, \citenamefont {Moreau},\
  and\ \citenamefont {{Ben Nessib}}}]{SAHALBRECHOT20141148}%
  \BibitemOpen
  \bibfield  {author} {\bibinfo {author} {\bibfnamefont {S.}~\bibnamefont
  {Sahal-Br{\'e}chot}}, \bibinfo {author} {\bibfnamefont {M.~S.}\ \bibnamefont
  {Dimitrijević}}, \bibinfo {author} {\bibfnamefont {N.}~\bibnamefont
  {Moreau}},\ and\ \bibinfo {author} {\bibfnamefont {N.}~\bibnamefont {{Ben
  Nessib}}},\ }\bibfield  {title} {\enquote {\bibinfo {title} {The {STARK-B}
  database as a resource for {``STARK''} widths and shifts data: State of
  advancement and program of development},}\ }\href
  {https://doi.org/https://doi.org/10.1016/j.asr.2013.08.015} {\bibfield
  {journal} {\bibinfo  {journal} {Advances in Space Research}\ }\textbf
  {\bibinfo {volume} {54}},\ \bibinfo {pages} {1148--1151} (\bibinfo {year}
  {2014})}\BibitemShut {NoStop}%
\bibitem [{\citenamefont {Alonso-Medina}(2019)}]{Alonso}%
  \BibitemOpen
  \bibfield  {author} {\bibinfo {author} {\bibfnamefont {A.}~\bibnamefont
  {Alonso-Medina}},\ }\bibfield  {title} {\enquote {\bibinfo {title}
  {Measurement of laser-induced plasma: {Stark} broadening parameters of {Pb II
  } 2203.5 and 4386.5 {$A^o$} spectral lines},}\ }\href
  {https://doi.org/10.1177/0003702818816305} {\bibfield  {journal} {\bibinfo
  {journal} {Applied Spectroscopy}\ }\textbf {\bibinfo {volume} {73}},\
  \bibinfo {pages} {133--151} (\bibinfo {year} {2019})},\ \bibinfo {note}
  {pMID: 30421963}\BibitemShut {NoStop}%
\bibitem [{\citenamefont {Ortiz}\ and\ \citenamefont
  {Mayo}(2005)}]{Ortiz_2005}%
  \BibitemOpen
  \bibfield  {author} {\bibinfo {author} {\bibfnamefont {M.}~\bibnamefont
  {Ortiz}}\ and\ \bibinfo {author} {\bibfnamefont {R.}~\bibnamefont {Mayo}},\
  }\bibfield  {title} {\enquote {\bibinfo {title} {Measurement of the {Stark}
  broadening for several lines of singly ionized gold},}\ }\href
  {https://doi.org/10.1088/0953-4075/38/22/003} {\bibfield  {journal} {\bibinfo
   {journal} {Journal of Physics B: Atomic, Molecular and Optical Physics}\
  }\textbf {\bibinfo {volume} {38}},\ \bibinfo {pages} {3953} (\bibinfo {year}
  {2005})}\BibitemShut {NoStop}%
\bibitem [{\citenamefont {Fleurier}, \citenamefont {Sahal-Brechot},\ and\
  \citenamefont {Chapelle}(1977)}]{Fleurier_1977}%
  \BibitemOpen
  \bibfield  {author} {\bibinfo {author} {\bibfnamefont {C.}~\bibnamefont
  {Fleurier}}, \bibinfo {author} {\bibfnamefont {S.}~\bibnamefont
  {Sahal-Brechot}},\ and\ \bibinfo {author} {\bibfnamefont {J.}~\bibnamefont
  {Chapelle}},\ }\bibfield  {title} {\enquote {\bibinfo {title} {{Stark}
  profiles of {Al~I} and {Al~II} lines},}\ }\href
  {https://doi.org/10.1088/0022-3700/10/17/018} {\bibfield  {journal} {\bibinfo
   {journal} {Journal of Physics B: Atomic and Molecular Physics}\ }\textbf
  {\bibinfo {volume} {10}},\ \bibinfo {pages} {3435} (\bibinfo {year}
  {1977})}\BibitemShut {NoStop}%
\bibitem [{\citenamefont {Cirisan}\ \emph {et~al.}(2014)\citenamefont
  {Cirisan}, \citenamefont {Cvejić}, \citenamefont {Gavrilović},
  \citenamefont {Jovićević}, \citenamefont {Konjević},\ and\ \citenamefont
  {Hermann}}]{CIRISAN2014652}%
  \BibitemOpen
  \bibfield  {author} {\bibinfo {author} {\bibfnamefont {M.}~\bibnamefont
  {Cirisan}}, \bibinfo {author} {\bibfnamefont {M.}~\bibnamefont {Cvejić}},
  \bibinfo {author} {\bibfnamefont {M.}~\bibnamefont {Gavrilović}}, \bibinfo
  {author} {\bibfnamefont {S.}~\bibnamefont {Jovićević}}, \bibinfo {author}
  {\bibfnamefont {N.}~\bibnamefont {Konjević}},\ and\ \bibinfo {author}
  {\bibfnamefont {J.}~\bibnamefont {Hermann}},\ }\bibfield  {title} {\enquote
  {\bibinfo {title} {{Stark} broadening measurement of {Al~II} lines in a
  laser-induced plasma},}\ }\href
  {https://doi.org/https://doi.org/10.1016/j.jqsrt.2013.10.002} {\bibfield
  {journal} {\bibinfo  {journal} {Journal of Quantitative Spectroscopy and
  Radiative Transfer}\ }\textbf {\bibinfo {volume} {133}},\ \bibinfo {pages}
  {652--662} (\bibinfo {year} {2014})}\BibitemShut {NoStop}%
\bibitem [{\citenamefont {Kielkopf}\ and\ \citenamefont
  {Allard}(2014)}]{Kielkopf_2014}%
  \BibitemOpen
  \bibfield  {author} {\bibinfo {author} {\bibfnamefont {J.~F.}\ \bibnamefont
  {Kielkopf}}\ and\ \bibinfo {author} {\bibfnamefont {N.~F.}\ \bibnamefont
  {Allard}},\ }\bibfield  {title} {\enquote {\bibinfo {title} {Shift and width
  of the {Balmer} series {H} alpha line at high electron density in a
  laser-produced plasma},}\ }\href
  {https://doi.org/10.1088/0953-4075/47/15/155701} {\bibfield  {journal}
  {\bibinfo  {journal} {Journal of Physics B: Atomic, Molecular and Optical
  Physics}\ }\textbf {\bibinfo {volume} {47}},\ \bibinfo {pages} {155701}
  (\bibinfo {year} {2014})}\BibitemShut {NoStop}%
\bibitem [{\citenamefont {Djurovi\ifmmode~\acute{c}\else \'{c}\fi{}}\ \emph
  {et~al.}(2009)\citenamefont {Djurovi\ifmmode~\acute{c}\else \'{c}\fi{}},
  \citenamefont {\ifmmode \acute{C}\else \'{C}\fi{}iri\ifmmode~\check{s}\else
  \v{s}\fi{}an}, \citenamefont {Demura}, \citenamefont {Demchenko},
  \citenamefont {Nikoli\ifmmode~\acute{c}\else \'{c}\fi{}}, \citenamefont
  {Gigosos},\ and\ \citenamefont {Gonz\'alez}}]{Djurovic}%
  \BibitemOpen
  \bibfield  {author} {\bibinfo {author} {\bibfnamefont {S.}~\bibnamefont
  {Djurovi\ifmmode~\acute{c}\else \'{c}\fi{}}}, \bibinfo {author}
  {\bibfnamefont {M.}~\bibnamefont {\ifmmode \acute{C}\else
  \'{C}\fi{}iri\ifmmode~\check{s}\else \v{s}\fi{}an}}, \bibinfo {author}
  {\bibfnamefont {A.~V.}\ \bibnamefont {Demura}}, \bibinfo {author}
  {\bibfnamefont {G.~V.}\ \bibnamefont {Demchenko}}, \bibinfo {author}
  {\bibfnamefont {D.}~\bibnamefont {Nikoli\ifmmode~\acute{c}\else \'{c}\fi{}}},
  \bibinfo {author} {\bibfnamefont {M.~A.}\ \bibnamefont {Gigosos}},\ and\
  \bibinfo {author} {\bibfnamefont {M.~A.}\ \bibnamefont {Gonz\'alez}},\
  }\bibfield  {title} {\enquote {\bibinfo {title} {Measurements of
  ${{H}}_{{\beta}}$ {Stark} central asymmetry and its analysis through standard
  theory and computer simulations},}\ }\href
  {https://doi.org/10.1103/PhysRevE.79.046402} {\bibfield  {journal} {\bibinfo
  {journal} {Phys. Rev. E}\ }\textbf {\bibinfo {volume} {79}},\ \bibinfo
  {pages} {046402} (\bibinfo {year} {2009})}\BibitemShut {NoStop}%
\bibitem [{\citenamefont {Gigosos}\ and\ \citenamefont
  {Gonz{\'a}lez}(2006)}]{Gigosos2006}%
  \BibitemOpen
  \bibfield  {author} {\bibinfo {author} {\bibfnamefont {M.~A.}\ \bibnamefont
  {Gigosos}}\ and\ \bibinfo {author} {\bibfnamefont {M.~{\'A}.}\ \bibnamefont
  {Gonz{\'a}lez}},\ }\bibfield  {title} {\enquote {\bibinfo {title} {Study on
  the asymmetry of the {Balmer} lines},}\ }\href
  {https://doi.org/10.1063/1.2406038} {\bibfield  {journal} {\bibinfo
  {journal} {AIP Conference Proceedings}\ }\textbf {\bibinfo {volume} {876}},\
  \bibinfo {pages} {294--300} (\bibinfo {year} {2006})}\BibitemShut {NoStop}%
\bibitem [{\citenamefont {Stambulchik}\ and\ \citenamefont
  {Maron}(2010)}]{STAMBULCHIK20109}%
  \BibitemOpen
  \bibfield  {author} {\bibinfo {author} {\bibfnamefont {E.}~\bibnamefont
  {Stambulchik}}\ and\ \bibinfo {author} {\bibfnamefont {Y.}~\bibnamefont
  {Maron}},\ }\bibfield  {title} {\enquote {\bibinfo {title} {Plasma line
  broadening and computer simulations: {A} mini-review},}\ }\href
  {https://doi.org/https://doi.org/10.1016/j.hedp.2009.07.001} {\bibfield
  {journal} {\bibinfo  {journal} {High Energy Density Physics}\ }\textbf
  {\bibinfo {volume} {6}},\ \bibinfo {pages} {9--14} (\bibinfo {year}
  {2010})}\BibitemShut {NoStop}%
\bibitem [{\citenamefont {{El Sherbini}}\ \emph {et~al.}(2005)\citenamefont
  {{El Sherbini}}, \citenamefont {{El Sherbini}}, \citenamefont {Hegazy},
  \citenamefont {Cristoforetti}, \citenamefont {Legnaioli}, \citenamefont
  {Palleschi}, \citenamefont {Pardini}, \citenamefont {Salvetti},\ and\
  \citenamefont {Tognoni}}]{ELSHERBINI20051573}%
  \BibitemOpen
  \bibfield  {author} {\bibinfo {author} {\bibfnamefont {A.}~\bibnamefont {{El
  Sherbini}}}, \bibinfo {author} {\bibfnamefont {T.}~\bibnamefont {{El
  Sherbini}}}, \bibinfo {author} {\bibfnamefont {H.}~\bibnamefont {Hegazy}},
  \bibinfo {author} {\bibfnamefont {G.}~\bibnamefont {Cristoforetti}}, \bibinfo
  {author} {\bibfnamefont {S.}~\bibnamefont {Legnaioli}}, \bibinfo {author}
  {\bibfnamefont {V.}~\bibnamefont {Palleschi}}, \bibinfo {author}
  {\bibfnamefont {L.}~\bibnamefont {Pardini}}, \bibinfo {author} {\bibfnamefont
  {A.}~\bibnamefont {Salvetti}},\ and\ \bibinfo {author} {\bibfnamefont
  {E.}~\bibnamefont {Tognoni}},\ }\bibfield  {title} {\enquote {\bibinfo
  {title} {Evaluation of self-absorption coefficients of aluminum emission
  lines in laser-induced breakdown spectroscopy measurements},}\ }\href
  {https://doi.org/https://doi.org/10.1016/j.sab.2005.10.011} {\bibfield
  {journal} {\bibinfo  {journal} {Spectrochimica Acta Part B: Atomic
  Spectroscopy}\ }\textbf {\bibinfo {volume} {60}},\ \bibinfo {pages}
  {1573--1579} (\bibinfo {year} {2005})}\BibitemShut {NoStop}%
\bibitem [{\citenamefont {Hermann}, \citenamefont {Boulmer-Leborgne},\ and\
  \citenamefont {Hong}(1998)}]{Hermann}%
  \BibitemOpen
  \bibfield  {author} {\bibinfo {author} {\bibfnamefont {J.}~\bibnamefont
  {Hermann}}, \bibinfo {author} {\bibfnamefont {C.}~\bibnamefont
  {Boulmer-Leborgne}},\ and\ \bibinfo {author} {\bibfnamefont {D.}~\bibnamefont
  {Hong}},\ }\bibfield  {title} {\enquote {\bibinfo {title} {Diagnostics of the
  early phase of an ultraviolet laser induced plasma by spectral line analysis
  considering self-absorption},}\ }\href {https://doi.org/10.1063/1.366639}
  {\bibfield  {journal} {\bibinfo  {journal} {Journal of Applied Physics}\
  }\textbf {\bibinfo {volume} {83}},\ \bibinfo {pages} {691--696} (\bibinfo
  {year} {1998})}\BibitemShut {NoStop}%
\bibitem [{\citenamefont {Harilal}\ \emph {et~al.}(2022)\citenamefont
  {Harilal}, \citenamefont {Phillips}, \citenamefont {Froula}, \citenamefont
  {Anoop}, \citenamefont {Issac},\ and\ \citenamefont {Beg}}]{HARILAL2018}%
  \BibitemOpen
  \bibfield  {author} {\bibinfo {author} {\bibfnamefont {S.~S.}\ \bibnamefont
  {Harilal}}, \bibinfo {author} {\bibfnamefont {M.~C.}\ \bibnamefont
  {Phillips}}, \bibinfo {author} {\bibfnamefont {D.~H.}\ \bibnamefont
  {Froula}}, \bibinfo {author} {\bibfnamefont {K.~K.}\ \bibnamefont {Anoop}},
  \bibinfo {author} {\bibfnamefont {R.~C.}\ \bibnamefont {Issac}},\ and\
  \bibinfo {author} {\bibfnamefont {F.~N.}\ \bibnamefont {Beg}},\ }\bibfield
  {title} {\enquote {\bibinfo {title} {Optical diagnostics of laser-produced
  plasmas},}\ }\href {https://doi.org/10.1103/RevModPhys.94.035002} {\bibfield
  {journal} {\bibinfo  {journal} {Rev. Mod. Phys.}\ }\textbf {\bibinfo {volume}
  {94}},\ \bibinfo {pages} {035002} (\bibinfo {year} {2022})}\BibitemShut
  {NoStop}%
\bibitem [{\citenamefont {Bengoechea}, \citenamefont {Aguilera},\ and\
  \citenamefont {Arag{\'o}n}(2006)}]{BENGOECHEA200669}%
  \BibitemOpen
  \bibfield  {author} {\bibinfo {author} {\bibfnamefont {J.}~\bibnamefont
  {Bengoechea}}, \bibinfo {author} {\bibfnamefont {J.}~\bibnamefont
  {Aguilera}},\ and\ \bibinfo {author} {\bibfnamefont {C.}~\bibnamefont
  {Arag{\'o}n}},\ }\bibfield  {title} {\enquote {\bibinfo {title} {Application
  of laser-induced plasma spectroscopy to the measurement of {Stark} broadening
  parameters},}\ }\href
  {https://doi.org/https://doi.org/10.1016/j.sab.2005.11.003} {\bibfield
  {journal} {\bibinfo  {journal} {Spectrochimica Acta Part B: Atomic
  Spectroscopy}\ }\textbf {\bibinfo {volume} {61}},\ \bibinfo {pages} {69--80}
  (\bibinfo {year} {2006})}\BibitemShut {NoStop}%
\bibitem [{\citenamefont {Burger}\ \emph {et~al.}(2019)\citenamefont {Burger},
  \citenamefont {Skrodzki}, \citenamefont {Jovanovic}, \citenamefont
  {Phillips},\ and\ \citenamefont {Harilal}}]{Burger2019}%
  \BibitemOpen
  \bibfield  {author} {\bibinfo {author} {\bibfnamefont {M.}~\bibnamefont
  {Burger}}, \bibinfo {author} {\bibfnamefont {P.~J.}\ \bibnamefont
  {Skrodzki}}, \bibinfo {author} {\bibfnamefont {I.}~\bibnamefont {Jovanovic}},
  \bibinfo {author} {\bibfnamefont {M.~C.}\ \bibnamefont {Phillips}},\ and\
  \bibinfo {author} {\bibfnamefont {S.~S.}\ \bibnamefont {Harilal}},\
  }\bibfield  {title} {\enquote {\bibinfo {title} {Laser-produced uranium
  plasma characterization and {Stark} broadening measurements},}\ }\href
  {https://doi.org/10.1063/1.5099643} {\bibfield  {journal} {\bibinfo
  {journal} {Physics of Plasmas}\ }\textbf {\bibinfo {volume} {26}},\ \bibinfo
  {pages} {093103} (\bibinfo {year} {2019})}\BibitemShut {NoStop}%
\bibitem [{\citenamefont {Gornushkin}\ \emph {et~al.}(1999)\citenamefont
  {Gornushkin}, \citenamefont {King}, \citenamefont {Smith}, \citenamefont
  {Omenetto},\ and\ \citenamefont {Winefordner}}]{GORNUSHKIN19991207}%
  \BibitemOpen
  \bibfield  {author} {\bibinfo {author} {\bibfnamefont {I.~B.}\ \bibnamefont
  {Gornushkin}}, \bibinfo {author} {\bibfnamefont {L.~A.}\ \bibnamefont
  {King}}, \bibinfo {author} {\bibfnamefont {B.~W.}\ \bibnamefont {Smith}},
  \bibinfo {author} {\bibfnamefont {N.}~\bibnamefont {Omenetto}},\ and\
  \bibinfo {author} {\bibfnamefont {J.~D.}\ \bibnamefont {Winefordner}},\
  }\bibfield  {title} {\enquote {\bibinfo {title} {Line broadening mechanisms
  in the low pressure laser-induced plasma},}\ }\href
  {https://doi.org/https://doi.org/10.1016/S0584-8547(99)00064-6} {\bibfield
  {journal} {\bibinfo  {journal} {Spectrochimica Acta Part B: Atomic
  Spectroscopy}\ }\textbf {\bibinfo {volume} {54}},\ \bibinfo {pages}
  {1207--1217} (\bibinfo {year} {1999})}\BibitemShut {NoStop}%
\bibitem [{\citenamefont {Dojić}\ \emph {et~al.}(2020)\citenamefont {Dojić},
  \citenamefont {Skočić}, \citenamefont {Bukvić},\ and\ \citenamefont
  {Djeniže}}]{densal3}%
  \BibitemOpen
  \bibfield  {author} {\bibinfo {author} {\bibfnamefont {D.}~\bibnamefont
  {Dojić}}, \bibinfo {author} {\bibfnamefont {M.}~\bibnamefont {Skočić}},
  \bibinfo {author} {\bibfnamefont {S.}~\bibnamefont {Bukvić}},\ and\ \bibinfo
  {author} {\bibfnamefont {S.}~\bibnamefont {Djeniže}},\ }\bibfield  {title}
  {\enquote {\bibinfo {title} {{Stark} broadening measurements of {Al~II},
  {Al~III} and {He~I} 388.86 nm spectral lines at high electron densities},}\
  }\href {https://doi.org/https://doi.org/10.1016/j.sab.2020.105816} {\bibfield
   {journal} {\bibinfo  {journal} {Spectrochimica Acta Part B: Atomic
  Spectroscopy}\ }\textbf {\bibinfo {volume} {166}},\ \bibinfo {pages} {105816}
  (\bibinfo {year} {2020})}\BibitemShut {NoStop}%
\bibitem [{\citenamefont {Konjević}\ and\ \citenamefont
  {Wiese}(1990)}]{density704}%
  \BibitemOpen
  \bibfield  {author} {\bibinfo {author} {\bibfnamefont {N.}~\bibnamefont
  {Konjević}}\ and\ \bibinfo {author} {\bibfnamefont {W.~L.}\ \bibnamefont
  {Wiese}},\ }\bibfield  {title} {\enquote {\bibinfo {title} {Experimental
  {Stark} widths and shifts for spectral lines of neutral and ionized atoms},}\
  }\href {https://doi.org/10.1063/1.555847} {\bibfield  {journal} {\bibinfo
  {journal} {Journal of Physical and Chemical Reference Data}\ }\textbf
  {\bibinfo {volume} {19}},\ \bibinfo {pages} {1307--1385} (\bibinfo {year}
  {1990})},\ \Eprint {https://arxiv.org/abs/https://doi.org/10.1063/1.555847}
  {https://doi.org/10.1063/1.555847} \BibitemShut {NoStop}%
\bibitem [{\citenamefont {Konjević}\ \emph {et~al.}(2002)\citenamefont
  {Konjević}, \citenamefont {Lesage}, \citenamefont {Fuhr},\ and\
  \citenamefont {Wiese}}]{density559}%
  \BibitemOpen
  \bibfield  {author} {\bibinfo {author} {\bibfnamefont {N.}~\bibnamefont
  {Konjević}}, \bibinfo {author} {\bibfnamefont {A.}~\bibnamefont {Lesage}},
  \bibinfo {author} {\bibfnamefont {J.~R.}\ \bibnamefont {Fuhr}},\ and\
  \bibinfo {author} {\bibfnamefont {W.~L.}\ \bibnamefont {Wiese}},\ }\bibfield
  {title} {\enquote {\bibinfo {title} {Experimental {Stark} widths and shifts
  for spectral lines of neutral and ionized atoms ({A} critical review of
  selected data for the period 1989 through 2000)},}\ }\href
  {https://doi.org/10.1063/1.1486456} {\bibfield  {journal} {\bibinfo
  {journal} {Journal of Physical and Chemical Reference Data}\ }\textbf
  {\bibinfo {volume} {31}},\ \bibinfo {pages} {819--927} (\bibinfo {year}
  {2002})},\ \Eprint {https://arxiv.org/abs/https://doi.org/10.1063/1.1486456}
  {https://doi.org/10.1063/1.1486456} \BibitemShut {NoStop}%
\bibitem [{\citenamefont {Allen}\ \emph {et~al.}(1975)\citenamefont {Allen},
  \citenamefont {Blaha}, \citenamefont {Jones}, \citenamefont {Sanchez},\ and\
  \citenamefont {Griem}}]{Allen}%
  \BibitemOpen
  \bibfield  {author} {\bibinfo {author} {\bibfnamefont {A.~W.}\ \bibnamefont
  {Allen}}, \bibinfo {author} {\bibfnamefont {M.}~\bibnamefont {Blaha}},
  \bibinfo {author} {\bibfnamefont {W.~W.}\ \bibnamefont {Jones}}, \bibinfo
  {author} {\bibfnamefont {A.}~\bibnamefont {Sanchez}},\ and\ \bibinfo {author}
  {\bibfnamefont {H.~R.}\ \bibnamefont {Griem}},\ }\bibfield  {title} {\enquote
  {\bibinfo {title} {{Stark}-broadening measurement and calculations for a
  singly ionized aluminum line},}\ }\href
  {https://doi.org/10.1103/PhysRevA.11.477} {\bibfield  {journal} {\bibinfo
  {journal} {Phys. Rev. A}\ }\textbf {\bibinfo {volume} {11}},\ \bibinfo
  {pages} {477--479} (\bibinfo {year} {1975})}\BibitemShut {NoStop}%
\bibitem [{\citenamefont {Heuschkel}\ and\ \citenamefont
  {Kusch}(1973)}]{heuschkel1973stark}%
  \BibitemOpen
  \bibfield  {author} {\bibinfo {author} {\bibfnamefont {J.}~\bibnamefont
  {Heuschkel}}\ and\ \bibinfo {author} {\bibfnamefont {H.}~\bibnamefont
  {Kusch}},\ }\bibfield  {title} {\enquote {\bibinfo {title} {{Stark}
  broadening and shift of singly ionized aluminium lines},}\ }\href@noop {}
  {\bibfield  {journal} {\bibinfo  {journal} {Astronomy and Astrophysics}\
  }\textbf {\bibinfo {volume} {25}},\ \bibinfo {pages} {149} (\bibinfo {year}
  {1973})}\BibitemShut {NoStop}%
\bibitem [{\citenamefont {Puri\ifmmode~\acute{c}\else \'{c}\fi{}},
  \citenamefont {\ifmmode~\acute{C}\else \'{C}\fi{}uk},\ and\ \citenamefont
  {Laki\ifmmode \acute{c}\else \'{c}\fi{}evi\ifmmode~\acute{c}\else
  \'{c}\fi{}}(1985)}]{Puri}%
  \BibitemOpen
  \bibfield  {author} {\bibinfo {author} {\bibfnamefont {J.}~\bibnamefont
  {Puri\ifmmode~\acute{c}\else \'{c}\fi{}}}, \bibinfo {author} {\bibfnamefont
  {M.}~\bibnamefont {\ifmmode~\acute{C}\else \'{C}\fi{}uk}},\ and\ \bibinfo
  {author} {\bibfnamefont {I.~S.}\ \bibnamefont {Laki\ifmmode \acute{c}\else
  \'{c}\fi{}evi\ifmmode~\acute{c}\else \'{c}\fi{}}},\ }\bibfield  {title}
  {\enquote {\bibinfo {title} {Regularities and systematic trends in the
  {Stark} broadening and shift parameters of spectral lines in plasma},}\
  }\href {https://doi.org/10.1103/PhysRevA.32.1106} {\bibfield  {journal}
  {\bibinfo  {journal} {Phys. Rev. A}\ }\textbf {\bibinfo {volume} {32}},\
  \bibinfo {pages} {1106--1114} (\bibinfo {year} {1985})}\BibitemShut {NoStop}%
\bibitem [{\citenamefont {Blagojević}\ and\ \citenamefont
  {Konjević}(2017)}]{BLAGOJEVIC20179}%
  \BibitemOpen
  \bibfield  {author} {\bibinfo {author} {\bibfnamefont {B.}~\bibnamefont
  {Blagojević}}\ and\ \bibinfo {author} {\bibfnamefont {N.}~\bibnamefont
  {Konjević}},\ }\bibfield  {title} {\enquote {\bibinfo {title} {Semiclassical
  calculations of electron impact {Stark} widths and shifts of singly ionized
  atom lines revisited},}\ }\href
  {https://doi.org/https://doi.org/10.1016/j.jqsrt.2017.04.025} {\bibfield
  {journal} {\bibinfo  {journal} {Journal of Quantitative Spectroscopy and
  Radiative Transfer}\ }\textbf {\bibinfo {volume} {198}},\ \bibinfo {pages}
  {9--24} (\bibinfo {year} {2017})}\BibitemShut {NoStop}%
\bibitem [{\citenamefont {Geethika}\ \emph {et~al.}(2023)\citenamefont
  {Geethika}, \citenamefont {Thomas}, \citenamefont {Patel}, \citenamefont
  {Kumar~R.},\ and\ \citenamefont {Joshi}}]{geethika_jaas}%
  \BibitemOpen
  \bibfield  {author} {\bibinfo {author} {\bibfnamefont {B.~R.}\ \bibnamefont
  {Geethika}}, \bibinfo {author} {\bibfnamefont {J.}~\bibnamefont {Thomas}},
  \bibinfo {author} {\bibfnamefont {M.}~\bibnamefont {Patel}}, \bibinfo
  {author} {\bibfnamefont {R.}~\bibnamefont {Kumar~R.}},\ and\ \bibinfo
  {author} {\bibfnamefont {H.~C.}\ \bibnamefont {Joshi}},\ }\bibfield  {title}
  {\enquote {\bibinfo {title} {Spatio-temporal dynamics of anisotropic emission
  from nano-second laser produced aluminium plasma},}\ }\href
  {https://doi.org/10.1039/D3JA00228D} {\bibfield  {journal} {\bibinfo
  {journal} {J. Anal. At. Spectrom.}\ }\textbf {\bibinfo {volume} {38}},\
  \bibinfo {pages} {2477--2485} (\bibinfo {year} {2023})}\BibitemShut {NoStop}%
\bibitem [{\citenamefont {Harilal}\ \emph {et~al.}(1997)\citenamefont
  {Harilal}, \citenamefont {Bindhu}, \citenamefont {Issac}, \citenamefont
  {Nampoori},\ and\ \citenamefont {Vallabhan}}]{densharilal}%
  \BibitemOpen
  \bibfield  {author} {\bibinfo {author} {\bibfnamefont {S.~S.}\ \bibnamefont
  {Harilal}}, \bibinfo {author} {\bibfnamefont {C.~V.}\ \bibnamefont {Bindhu}},
  \bibinfo {author} {\bibfnamefont {R.~C.}\ \bibnamefont {Issac}}, \bibinfo
  {author} {\bibfnamefont {V.~P.~N.}\ \bibnamefont {Nampoori}},\ and\ \bibinfo
  {author} {\bibfnamefont {C.~P.~G.}\ \bibnamefont {Vallabhan}},\ }\bibfield
  {title} {\enquote {\bibinfo {title} {Electron density and temperature
  measurements in a laser produced carbon plasma},}\ }\href
  {https://doi.org/10.1063/1.366276} {\bibfield  {journal} {\bibinfo  {journal}
  {Journal of Applied Physics}\ }\textbf {\bibinfo {volume} {82}},\ \bibinfo
  {pages} {2140--2146} (\bibinfo {year} {1997})},\ \Eprint
  {https://arxiv.org/abs/https://doi.org/10.1063/1.366276}
  {https://doi.org/10.1063/1.366276} \BibitemShut {NoStop}%
\bibitem [{\citenamefont {Arag{\'o}n}\ and\ \citenamefont
  {Aguilera}(2008)}]{ARAGON2008893}%
  \BibitemOpen
  \bibfield  {author} {\bibinfo {author} {\bibfnamefont {C.}~\bibnamefont
  {Arag{\'o}n}}\ and\ \bibinfo {author} {\bibfnamefont {J.}~\bibnamefont
  {Aguilera}},\ }\bibfield  {title} {\enquote {\bibinfo {title}
  {Characterization of laser induced plasmas by optical emission spectroscopy:
  {A} review of experiments and methods},}\ }\href
  {https://doi.org/https://doi.org/10.1016/j.sab.2008.05.010} {\bibfield
  {journal} {\bibinfo  {journal} {Spectrochimica Acta Part B: Atomic
  Spectroscopy}\ }\textbf {\bibinfo {volume} {63}},\ \bibinfo {pages}
  {893--916} (\bibinfo {year} {2008})}\BibitemShut {NoStop}%
\bibitem [{\citenamefont {Cristoforetti}\ \emph {et~al.}(2010)\citenamefont
  {Cristoforetti}, \citenamefont {{De Giacomo}}, \citenamefont {Dell'Aglio},
  \citenamefont {Legnaioli}, \citenamefont {Tognoni}, \citenamefont
  {Palleschi},\ and\ \citenamefont {Omenetto}}]{CRISTOFORETTI201086}%
  \BibitemOpen
  \bibfield  {author} {\bibinfo {author} {\bibfnamefont {G.}~\bibnamefont
  {Cristoforetti}}, \bibinfo {author} {\bibfnamefont {A.}~\bibnamefont {{De
  Giacomo}}}, \bibinfo {author} {\bibfnamefont {M.}~\bibnamefont {Dell'Aglio}},
  \bibinfo {author} {\bibfnamefont {S.}~\bibnamefont {Legnaioli}}, \bibinfo
  {author} {\bibfnamefont {E.}~\bibnamefont {Tognoni}}, \bibinfo {author}
  {\bibfnamefont {V.}~\bibnamefont {Palleschi}},\ and\ \bibinfo {author}
  {\bibfnamefont {N.}~\bibnamefont {Omenetto}},\ }\bibfield  {title} {\enquote
  {\bibinfo {title} {Local thermodynamic equilibrium in laser-induced breakdown
  spectroscopy: Beyond the {McWhirter} criterion},}\ }\href
  {https://doi.org/https://doi.org/10.1016/j.sab.2009.11.005} {\bibfield
  {journal} {\bibinfo  {journal} {Spectrochimica Acta Part B: Atomic
  Spectroscopy}\ }\textbf {\bibinfo {volume} {65}},\ \bibinfo {pages} {86--95}
  (\bibinfo {year} {2010})}\BibitemShut {NoStop}%
\bibitem [{\citenamefont {Kramida}\ \emph {et~al.}(2018)\citenamefont
  {Kramida}, \citenamefont {Ralchenko}, \citenamefont {Reader},\ and\
  \citenamefont {{NIST ASD Team}}}]{NIST_ASD}%
  \BibitemOpen
  \bibfield  {author} {\bibinfo {author} {\bibfnamefont {A.}~\bibnamefont
  {Kramida}}, \bibinfo {author} {\bibfnamefont {Y.}~\bibnamefont {Ralchenko}},
  \bibinfo {author} {\bibfnamefont {J.}~\bibnamefont {Reader}},\ and\ \bibinfo
  {author} {\bibnamefont {{NIST ASD Team}}},\ }\href
  {https://doi.org/10.18434/T4W30F} {\enquote {\bibinfo {title} {{NIST} atomic
  spectra database (ver. 5.5.6)},}\ }\bibinfo {howpublished} {[Online].
  Available: \url{https://physics.nist.gov/asd}} (\bibinfo {year} {2018}),\
  \bibinfo {note} {national Institute of Standards and Technology,
  Gaithersburg, MD. Accessed: 2018, July 4}\BibitemShut {NoStop}%
\bibitem [{\citenamefont {Zehra}\ \emph {et~al.}(2017)\citenamefont {Zehra},
  \citenamefont {Bashir}, \citenamefont {Hassan}, \citenamefont {Ahmed},
  \citenamefont {Akram},\ and\ \citenamefont
  {Hayat}}]{Zehra_Bashir_Hassan_Ahmed_Akram_Hayat_2017}%
  \BibitemOpen
  \bibfield  {author} {\bibinfo {author} {\bibfnamefont {K.}~\bibnamefont
  {Zehra}}, \bibinfo {author} {\bibfnamefont {S.}~\bibnamefont {Bashir}},
  \bibinfo {author} {\bibfnamefont {S.}~\bibnamefont {Hassan}}, \bibinfo
  {author} {\bibfnamefont {Q.}~\bibnamefont {Ahmed}}, \bibinfo {author}
  {\bibfnamefont {M.}~\bibnamefont {Akram}},\ and\ \bibinfo {author}
  {\bibfnamefont {A.}~\bibnamefont {Hayat}},\ }\bibfield  {title} {\enquote
  {\bibinfo {title} {The effect of nature and pressure of ambient environment
  on laser-induced breakdown spectroscopy and ablation mechanisms of {Si}},}\
  }\href {https://doi.org/10.1017/S0263034617000477} {\bibfield  {journal}
  {\bibinfo  {journal} {Laser and Particle Beams}\ }\textbf {\bibinfo {volume}
  {35}},\ \bibinfo {pages} {492--504} (\bibinfo {year} {2017})}\BibitemShut
  {NoStop}%
\bibitem [{\citenamefont {Kautz}\ \emph {et~al.}(2020)\citenamefont {Kautz},
  \citenamefont {Yeak}, \citenamefont {Bernacki}, \citenamefont {Phillips},\
  and\ \citenamefont {Harilal}}]{Kautz2020}%
  \BibitemOpen
  \bibfield  {author} {\bibinfo {author} {\bibfnamefont {E.~J.}\ \bibnamefont
  {Kautz}}, \bibinfo {author} {\bibfnamefont {J.}~\bibnamefont {Yeak}},
  \bibinfo {author} {\bibfnamefont {B.~E.}\ \bibnamefont {Bernacki}}, \bibinfo
  {author} {\bibfnamefont {M.~C.}\ \bibnamefont {Phillips}},\ and\ \bibinfo
  {author} {\bibfnamefont {S.~S.}\ \bibnamefont {Harilal}},\ }\bibfield
  {title} {\enquote {\bibinfo {title} {The role of ambient gas confinement,
  plasma chemistry, and focusing conditions on emission features of femtosecond
  laser-produced plasmas},}\ }\href {https://doi.org/10.1039/D0JA00111B}
  {\bibfield  {journal} {\bibinfo  {journal} {Journal of Analytical Atomic
  Spectrometry}\ }\textbf {\bibinfo {volume} {35}},\ \bibinfo {pages}
  {1574--1586} (\bibinfo {year} {2020})}\BibitemShut {NoStop}%
\bibitem [{\citenamefont {Harilal}\ \emph {et~al.}(2003)\citenamefont
  {Harilal}, \citenamefont {Bindhu}, \citenamefont {Tillack}, \citenamefont
  {Najmabadi},\ and\ \citenamefont {Gaeris}}]{Harilal2003}%
  \BibitemOpen
  \bibfield  {author} {\bibinfo {author} {\bibfnamefont {S.~S.}\ \bibnamefont
  {Harilal}}, \bibinfo {author} {\bibfnamefont {C.~V.}\ \bibnamefont {Bindhu}},
  \bibinfo {author} {\bibfnamefont {M.~S.}\ \bibnamefont {Tillack}}, \bibinfo
  {author} {\bibfnamefont {F.}~\bibnamefont {Najmabadi}},\ and\ \bibinfo
  {author} {\bibfnamefont {A.~C.}\ \bibnamefont {Gaeris}},\ }\bibfield  {title}
  {\enquote {\bibinfo {title} {Internal structure and expansion dynamics of
  laser ablation plumes into ambient gases},}\ }\href
  {https://doi.org/10.1063/1.1544070} {\bibfield  {journal} {\bibinfo
  {journal} {Journal of Applied Physics}\ }\textbf {\bibinfo {volume} {93}},\
  \bibinfo {pages} {2380--2388} (\bibinfo {year} {2003})}\BibitemShut {NoStop}%
\bibitem [{\citenamefont {Joyce}, \citenamefont {Woltz},\ and\ \citenamefont
  {Hooper}(1987)}]{joyce_asym}%
  \BibitemOpen
  \bibfield  {author} {\bibinfo {author} {\bibfnamefont {R.~F.}\ \bibnamefont
  {Joyce}}, \bibinfo {author} {\bibfnamefont {L.~A.}\ \bibnamefont {Woltz}},\
  and\ \bibinfo {author} {\bibfnamefont {C.~F.}\ \bibnamefont {Hooper}},\
  }\bibfield  {title} {\enquote {\bibinfo {title} {Asymmetry of
  {Stark}-broadened {Lyman} lines from laser-produced plasmas},}\ }\href
  {https://doi.org/10.1103/PhysRevA.35.2228} {\bibfield  {journal} {\bibinfo
  {journal} {Phys. Rev. A}\ }\textbf {\bibinfo {volume} {35}},\ \bibinfo
  {pages} {2228--2233} (\bibinfo {year} {1987})}\BibitemShut {NoStop}%
\bibitem [{\citenamefont {Thomas}\ \emph {et~al.}(2020)\citenamefont {Thomas},
  \citenamefont {Joshi}, \citenamefont {Kumar},\ and\ \citenamefont
  {Philip}}]{PhysRevE_jinto}%
  \BibitemOpen
  \bibfield  {author} {\bibinfo {author} {\bibfnamefont {J.}~\bibnamefont
  {Thomas}}, \bibinfo {author} {\bibfnamefont {H.~C.}\ \bibnamefont {Joshi}},
  \bibinfo {author} {\bibfnamefont {A.}~\bibnamefont {Kumar}},\ and\ \bibinfo
  {author} {\bibfnamefont {R.}~\bibnamefont {Philip}},\ }\bibfield  {title}
  {\enquote {\bibinfo {title} {Observation of ion acceleration in nanosecond
  laser generated plasma on a nickel thin film under rear ablation geometry},}\
  }\href {https://doi.org/10.1103/PhysRevE.102.043205} {\bibfield  {journal}
  {\bibinfo  {journal} {Phys. Rev. E}\ }\textbf {\bibinfo {volume} {102}},\
  \bibinfo {pages} {043205} (\bibinfo {year} {2020})}\BibitemShut {NoStop}%
\bibitem [{\citenamefont {Ferri}\ \emph {et~al.}(2014)\citenamefont {Ferri},
  \citenamefont {Calisti}, \citenamefont {Moss{\'e}}, \citenamefont {Rosato},
  \citenamefont {Talin}, \citenamefont {Alexiou}, \citenamefont {Gigosos},
  \citenamefont {Gonz{\'a}lez}, \citenamefont {Gonz{\'a}lez-Herrero},
  \citenamefont {Lara}, \citenamefont {Gomez}, \citenamefont {Iglesias},
  \citenamefont {Lorenzen}, \citenamefont {Mancini},\ and\ \citenamefont
  {Stambulchik}}]{ion_dynamics}%
  \BibitemOpen
  \bibfield  {author} {\bibinfo {author} {\bibfnamefont {S.}~\bibnamefont
  {Ferri}}, \bibinfo {author} {\bibfnamefont {A.}~\bibnamefont {Calisti}},
  \bibinfo {author} {\bibfnamefont {C.}~\bibnamefont {Moss{\'e}}}, \bibinfo
  {author} {\bibfnamefont {J.}~\bibnamefont {Rosato}}, \bibinfo {author}
  {\bibfnamefont {B.}~\bibnamefont {Talin}}, \bibinfo {author} {\bibfnamefont
  {S.}~\bibnamefont {Alexiou}}, \bibinfo {author} {\bibfnamefont {M.~A.}\
  \bibnamefont {Gigosos}}, \bibinfo {author} {\bibfnamefont {M.~A.}\
  \bibnamefont {Gonz{\'a}lez}}, \bibinfo {author} {\bibfnamefont
  {D.}~\bibnamefont {Gonz{\'a}lez-Herrero}}, \bibinfo {author} {\bibfnamefont
  {N.}~\bibnamefont {Lara}}, \bibinfo {author} {\bibfnamefont {T.}~\bibnamefont
  {Gomez}}, \bibinfo {author} {\bibfnamefont {C.}~\bibnamefont {Iglesias}},
  \bibinfo {author} {\bibfnamefont {S.}~\bibnamefont {Lorenzen}}, \bibinfo
  {author} {\bibfnamefont {R.~C.}\ \bibnamefont {Mancini}},\ and\ \bibinfo
  {author} {\bibfnamefont {E.}~\bibnamefont {Stambulchik}},\ }\bibfield
  {title} {\enquote {\bibinfo {title} {Ion dynamics effect on {Stark}-broadened
  line shapes: {A} cross-comparison of various models},}\ }\href
  {https://doi.org/10.3390/atoms2030299} {\bibfield  {journal} {\bibinfo
  {journal} {Atoms}\ }\textbf {\bibinfo {volume} {2}},\ \bibinfo {pages}
  {299--318} (\bibinfo {year} {2014})}\BibitemShut {NoStop}%
\bibitem [{\citenamefont {Hermann}\ \emph {et~al.}(2017)\citenamefont
  {Hermann}, \citenamefont {Grojo}, \citenamefont {Axente}, \citenamefont
  {Gerhard}, \citenamefont {Burger},\ and\ \citenamefont
  {Craciun}}]{Hermann2017}%
  \BibitemOpen
  \bibfield  {author} {\bibinfo {author} {\bibfnamefont {J.}~\bibnamefont
  {Hermann}}, \bibinfo {author} {\bibfnamefont {D.}~\bibnamefont {Grojo}},
  \bibinfo {author} {\bibfnamefont {E.}~\bibnamefont {Axente}}, \bibinfo
  {author} {\bibfnamefont {C.}~\bibnamefont {Gerhard}}, \bibinfo {author}
  {\bibfnamefont {M.~c.~v.}\ \bibnamefont {Burger}},\ and\ \bibinfo {author}
  {\bibfnamefont {V.}~\bibnamefont {Craciun}},\ }\bibfield  {title} {\enquote
  {\bibinfo {title} {Ideal radiation source for plasma spectroscopy generated
  by laser ablation},}\ }\href {https://doi.org/10.1103/PhysRevE.96.053210}
  {\bibfield  {journal} {\bibinfo  {journal} {Phys. Rev. E}\ }\textbf {\bibinfo
  {volume} {96}},\ \bibinfo {pages} {053210} (\bibinfo {year}
  {2017})}\BibitemShut {NoStop}%
\end{thebibliography}
\end{document}